
\magnification=1200
\hsize 15true cm \hoffset=0.5true cm
\vsize 23true cm
\baselineskip=15pt

\font\grande=cmr10 scaled \magstep4
\font\medio=cmr10 scaled \magstep2
\outer\def\beginsection#1\par{\medbreak\bigskip
      \message{#1}\leftline{\bf#1}\nobreak\medskip\vskip-\parskip
      \noindent}

\def \me {\buildrel <\over \sim}
\def \Me {\buildrel >\over \sim}
\def \pa {\partial}
\def \ra {\rightarrow}
\def \big {\bigtriangledown}
\def \fb {\overline \phi}
\def \rb {\overline \rho}

\def \pr {\prime}
\def \se {{\prime \prime}}
\def \H {{a^\prime \over a}}
\def \fp {{\phi^\prime}}

\def \la {\lambda}
\def \La {\Lambda}
\def \Da {\Delta}
\def \b {\beta}
\def \a {\alpha}
\def \ap {\alpha^\prime}

\def \Ga {\Gamma}
\def \ga {\gamma}
\def \sg {\sigma}
\def \da {\delta}
\def \ep {\epsilon}
\def \r {\rho}
\def \c {\chi}
\def \om {\omega}
\def \Om {\Omega}
\def \noi {\noindent}

\def\sqr#1#2{{\vcenter{\hrule height.#2pt\hbox{\vrule width.#2pt
height#1pt \kern#1pt\vrule width.#2pt}\hrule height.#2pt}}}

\def\lsim{\mathrel{\rlap{\lower4pt\hbox{\hskip1pt$\sim$}}
    \raise1pt\hbox{$<$}}}         
\def\gsim{\mathrel{\rlap{\lower4pt\hbox{\hskip1pt$\sim$}}
    \raise1pt\hbox{$>$}}}         

\nopagenumbers
\line{\hfil CERN-TH.7178/94}
\vskip 2 cm
\centerline {\grande  Dilaton Production}
\vskip 0.5 true cm
\centerline{\grande in String Cosmology}
\vskip 1true cm
\centerline{M.Gasperini}
\centerline{\it Dipartimento di Fisica Teorica, Via P.Giuria 1, 10125 Turin,
Italy,}
\centerline{\it and INFN, Sezione di Torino, Turin, Italy}
\centerline{and}
\centerline{G.Veneziano}
\centerline{\it Theory Division, CERN, Geneva, Switzerland}

\vskip 1.5 true cm
\centerline{\medio Abstract}

\noindent
We consider the coupled
evolution of density, (scalar) metric and dilaton perturbations
  in the transition from a
``stringy" phase of growing curvature and gravitational coupling to the
standard radiation-dominated cosmology.
We show that dilaton production, with a spectrum
tilted towards large frequencies, emerges as a general property of
this scenario.
We discuss the frame-independence of the dilaton spectrum and of the
inflationary properties of the metric background by using, as model of
source, a pressureless gas of weakly interacting strings, which is shown
to provide an approximate but consistent solution to the full system of
background equations and string equations of motion.
We combine various cosmological
bounds on a growing dilaton spectrum with the bound on the dilaton mass
obtained from tests of the equivalence principle, and we find allowed
windows compatible with a universe presently dominated
by a relic background of dilatonic dark matter.

\vskip 1 cm
\noindent
CERN-TH.7178/94

\noindent
February 1994
\vfill\eject

\footline={\hss\rm\folio\hss}
\pageno=1

\centerline{\bf 1. Introduction and motivations.}

It is well known that fluctuations of the metric background are
amplified in the context of inflationary cosmologies, and that the
amplification of their
transverse, traceless (spin $2$) component can be interpreted
as graviton production [1]. Models of the early
universe based on the low-energy string effective action (what we shall
refer to, for short, as ``string  cosmology"), are characterized by the
additional presence  of a dilaton background, $\phi (t)$.
 It is natural to expect
   an  amplification of the fluctuations
  $\da \phi \equiv \chi$ of the dilaton background -- with corresponding
dilaton production -- to accompany that of the metric
for a suitable time-evolution of the cosmological fields.

In this paper we discuss such a dilaton production in the context of the
so-called ``pre-big-bang' scenario [2], characterized by an accelerated
evolution from a flat, cold and weakly coupled initial regime to a final
hot, highly curved, strong coupling regime,   marking the beginning of
the standard ``post-big-bang" decelerated FRW cosmology. With this
background, the spectrum of the produced dilatons tends to grow with
frequency, just like in the (previously discussed [2,3]) case of graviton
production. As we shall see in Sec.3, the high frequency part of the
spectral distribution, $\Om_\chi$, of the dilaton energy density can be
parameterized (in units of critical density $\rho_c=H^2/G = M_p^2 H^2$) as
$$
\Om_\chi (\om,t)={\om \over \rho_c}{d\r_\chi\over d \om}\simeq
GH_1^2\left (\om\over \om_1\right)^\da \left(H_1\over H\right)^2
\left(a_1\over a\right)^4,
\,\,\,\,\,\,\,\, \da >0, \,\,\,\,\,\,\,\, \om < \om_1 \eqno(1.1)
$$
 where $M_p \simeq 10^{19}$ GeV is the Planck mass.
Here $H_1$ is the curvature scale evaluated at the time $t_1$ marking the
end of the inflationary epoch (which we assume to coincide with the
beginning of radiation dominance); $\om_1=H_1 a_1/a$ is the
maximum amplified proper frequency; $H=\dot a/a$, where $a$, as usual, is
the scale factor of the background metric. The integral over $\om$
is thus
dominated by the highest frequency $\om_1$,
$$
\Om_\chi (t)=\int ^{\om_1} {d\om\over \om}\Om_\chi (\om,t)
\simeq GH_1^2 \left(H_1\over H\right)^2\left(a_1\over a\right)^4
\eqno(1.2)
$$
and the condition $\Om_\chi<1$, required to avoid that the dilatons
overclose the universe in the radiation-dominated era ($a\sim H^{-1/2}$),
yields the constraint $H_1\me M_p$, already known [2,3] from the
graviton spectrum.

The produced dilatons, however, cannot be massless. Large distance
dilaton couplings can be estimated [4] and turn out to be at least of
 gravitational strength. This violates the universality of gravity
at low energy and, in particular,   induces corrections to the effective Newton
potential (in the static, weak-field limit), which may be reconciled with the
present tests of the equivalence principle [5] only for
 a dilaton mass satisfying [4,6]:
$$
m\Me m_0 \equiv 10^{-4}\, eV \eqno(1.3)
$$
 The expression (1.2) is thus valid only until the energy
density stays dominated by the relativistic modes, with
$\om(t)>m$. But, at the present time $t_0$ (with $H_0 \sim
10^{-61}\,M_p$), the maximum frequency $\om_1$ is
$$
\om_1(t_0)=H_1 {a_1\over a_0} \simeq  10^{-4}
\left(H_1\over M_p\right)^{1/2}\,eV
 \eqno(1.4)
$$
As $H_1<M_p$, even the highest mode must then become non-relativistic
before the present epoch, because of the
constraint (1.3). At the scale $H(t_m)=m$ the modes with $\om(t_m)\me m$
begin to oscillate coherently, with frequency $m$, and when they are
dominant the dilaton energy density becomes non-relativistic, with
$$
\Om_\chi (t)
\simeq GmH_1 \left(H_1\over H\right)^2 \left(m\over H_1\right)^{\da -1
\over 2}
\left(a_1\over a\right)^3, ~~~~~~~~~~ 0\leq \da \leq 1
$$
$$
\Om_\chi (t)
\simeq GmH_1 \left(H_1\over H\right)^2\left(a_1\over a\right)^3,
{}~~~~~~~~~~~~~~~~ \da \geq 1
\eqno(1.5)
$$
(the dependence on the spectral index $\da$ disappears for fast enough
growth of the spectrum, as shown in Sec.5).
Such a fraction of critical density grows in time during the radiation era,
while in the matter era ($a\sim H^{-2/3}$) it becomes fixed at the maximum
constant value
$$
\Om_\chi \simeq G m^2\left(H_1\over H_2\right)^{1/2}
\left(m\over H_1\right)^{\da -3 \over 2}, ~~~~~~~~~~ 0\geq \da \geq 1
$$
$$
\Om_\chi
\simeq GmH_1 \left(H_1\over H_2\right)^{1/2},~~~~~~~~~~~~ \da \geq 1
\eqno(1.6)
$$
where $H_2 \sim 10^{6}\,H_0$ is the curvature scale at the
matter-radiation transition.  The condition $\Om_\c\leq 1$ provides then,
for any
given inflation scale $H_1$, an upper limit for the dilaton mass,
$$
m \me \left(H_2 M_p^4H_1^{\da-4} \right)^{1/(\da+1)},
{}~~~~~~~~~~~~~~~~~ 0\leq \da \leq 1
$$
$$
m \me H_2^{1/2} M_p^2H_1^{-3/2},
{}~~~~~~~~~~~~~~~~~~~~~~~~~~~  \da \geq 1 \eqno(1.7)
$$
valid
for $m<H_1$ under the assumption that its life-time
is sufficiently long to reach the matter-dominated era
(if $m>H_1$ the dilaton must decay before
becoming dominant with respect to the
radiation, as we shall see in Sec.5, and the critical density
bound cannot be applied).

In addition to the constraint (1.7), which is an unavoidable consequence
of the quantum fluctuations of the dilaton background, one should also
consider, in general, the constraints following from  possible classical
oscillations of the background  around the minimum of the potential [7]. The
initial amplitude of such oscillations is, however, to a large extent
model-dependent and, as
discussed in Sec.5, we shall work under the assumption that
classical oscillations are negligible with respect to the quantum
fluctuations amplified by the cosmological evolution. This assumption will
give us the {\it maximum}
allowed region in parameter space.

In the absence of classical oscillations, the upper limit on $m$ obtained
from eq.(1.7), combined with the
lower limit (1.3), define for each value of $\da$ and
$H_1$ an allowed window for the dilaton mass such that, near the upper
end of the window, the produced dilatons can close the universe. Such a
dilaton dominance, however, can last only until
their energy density is dissipated into radiation,
which occurs
at a decay scale
$$
H_d \simeq \Gamma_d \simeq {m^3\over M_p^2} \eqno(1.8)
$$
We are thus led to the first interesting result of this analysis. If
$\da \Me 0.1$, the upper limit (1.7) turns out to be larger than the
lower bound (1.3) even for inflation scales $H_1\geq 10^{-5}~M_p$.
Moreover, always for $H_1\geq 10^{-5}~M_p$, the largest value of $m$
allowed by eq.(1.7) is $m\simeq 100~MeV$ (obtained for $\da \geq 1$),
and it implies $H_d \me H_0$. This means that, for fast enough growing
 spectra, and ``realistic" (at least in a string cosmology
context) inflation scales $H_1\geq 10^{-5}~M_p$, we can be
left today with
a background of relic dilatons possibly representing
 a significant fraction of the
dark matter background [8]. The allowed ranges of $m$
corresponding to this interesting
 possibility lie around the upper limits given in (1.7) and thus
depend on
  $\da$ and             $H_1$   in a complicated way.
   For $0.1 \me \da \me 0.72$ the range of $H_1$ for which this
possibility can be realized is given by
$$
H_1                  \me 10^{-(23-32\da)/(4-\da)} M_p
\eqno(1.9)
$$
while for $\da > 0.72$ values of $H_1/M_p$ up to $1$ are possible
(the case $\da \geq 1$ is illustrated in {\bf Fig.1}). The lower bound
on $\da$ is imposed by the simultaneous requirements $H_1
\geq 10^{-5}~M_p$ and $m\geq m_0$, together with eq.(1.7).
As $H_1/M_p$ is varied between $10^{-5}$ and $1$ the corresponding
dilaton mass vary over the whole domain from $10^{-4}~eV$ to $100~ MeV$.
 For lower
spectral slopes ($\da <0.1$), the present existence of a dominant dilaton
background becomes possible only for (unrealistically) low inflation
scales, as first pointed out in Ref.[9]
for the case of scalar perturbations
with a flat ($\da =0$) spectrum. Note that, according to eq.(1.9), a
final inflation scale $H_1$ exactly coinciding with the string scale
itself, $\sim 10^{-1}~M_p$, would be compatible with a present light
dilaton dominance only for $\da \Me 0.61$.

The inflation scale $H_1$ determines also the amplitude
of  scalar perturbations $\psi$ of the metric itself, and it is thus
constrained by the scalar contribution to the  CMBR anisotropies (the
contribution of tensor perturbations turns out to be negligible in our
context, as their spectrum grows very fast with frequency [2,3]).  The
behavior of the scalar perturbation spectrum, as we shall see,
depends in general on the adopted model of matter sources and background
evolution, and it is fair to say that our present
 knowledge of the details of
the stringy pre-big-bang
phase is to poor to make stringent predictions on
the exact value of the spectral index $\da$. On the other hand,
fortunately enough, the properties of a massive dilaton background are
only weakly dependent on the value of the spectral index for $\da >0$,
and rapidly become spectrum-independent as soon as $\da \geq 1$.
The particular example chosen in this
paper (see Sec.4) to discuss dilaton
 production, namely a three-dimensional,
isotropic, dilaton-dominated background with negligible matter sources,
gives the same spectrum (very fast growing, $\da =3$),
for scalar ($\psi$), dilaton ($\c$) and tensor
($h_{\mu\nu}$) perturbations. However, such an example is chosen here
for simplicity reasons only, in order
 to develop a first qualitative sketch
of  the scenario associated with dilaton production, and it should not
be taken as
particularly indicative for what regards the spectral properties of the
metric perturbations.
For a phenomenological discussion it is better to leave open
the possibility of different spectra
 for $\psi$ and $\c$ (possibility which is
in general allowed in this context, as we shall see in Sec.3), and to
parameterize the scalar (metric) energy density as
$$
\Om_\psi (\om,t) \simeq
GH_1^2\left(\om\over \om_1\right)^{n-1} \left(H_1\over H\right)^2
\left(a_1\over a \right)^4
\eqno(1.10)
$$
with $n-1$ in general different from $\da$.

The interesting
question to ask at this point is whether,
                 in the same range
of $H_1$ which we believe to be realistic, it is possible to produce
enough dilatons to close the present universe and, at the same time, to
generate scalar perturbations with a spectrum consistent with the
anisotropy observed by COBE [10]. This amounts to require
$$
10^{-5} \me {H_1\over M_p}\me 1,~~~~~~
\Om_\c (t_0) \simeq 1, ~~~~~~
\Om_\psi (\om_0,t_2) \simeq
GH_1^2 \left(\om_0\over \om_1\right)^{n-1} \simeq 10^{-10} \eqno(1.11)
$$
where $\om_0=H_0$ is the minimum amplified frequency corresponding to
a wave crossing today the Hubble radius, and $t_2$ is the time of
matter-radiation equilibrium,  nearly coincident with the
time of recombination.

The answer, perhaps surprisingly, is yes: the last requirement of
eq.(1.11) is
compatible with eq.(1.9), for the same range of $H_1$,
provided the scalar spectrum is also growing,
with
$$
1 \me n \me 1.34 \eqno(1.12)
$$
This allowed range of $n$
is well contained in the range of the
spectral index originally determined by COBE [10], $n=1\pm 0.5$,  and is
also consistent with the new recent fit [11] which gives $n=1.5\pm 0.5$. It
may be interesting to recall, in this context, that growing scalar
perturbations, with $n\simeq 1.25$, are also required for a simultaneous
fit of the COBE anisotropies and of the observed bulk motion and large
voids structures on a $50\, Mpc$ scale [12]. Growing scalar spectra can be
obtained in the ``hybrid inflation" model proposed by Linde [13] and
recently generalized to the class of ``false vacuum inflation" [14]
(see also [15]). Note also that the condition (1.12) would not be
incompatible with the lower bound on $\da$ required for a present
dominant dilaton background (according to eq.(1.9)), even in the case of
equal scalar and dilaton spectrum, $\da=n-1$.

Concluding this qualitative analysis, we can say  that the
possibility of producing a dilaton background which saturates the closure
density, together with scalar perturbations which provide the
observed cosmic anisotropies, seems to be naturally associated with a
growing  dilaton spectrum, $\da >0$. The fact that such a spectrum is
typical of string-based  pre-big-bang models represents, in our opinion, an
 interesting
aspect of such models, and motivates the study of dilaton production in
the string cosmology scenario. A requirement analogous to eq.(1.11),
formulated however in the context of extended inflation models where the
fluctuation spectrum of the Brans-Dicke scalar is not growing, may be
satisfied [16] only for a reheating temperature $T_r<10^{13}\,GeV$, namely
for very low scales $H_1\simeq T_r^2 /M_p\le 10^{-12}\,M_p$. We note,
finally, that the possibility of inflationary production of massless scalar
particles, associated with excitations of the Brans-Dicke field, was also
pointed out in Ref.[17], and previously discussed in Ref.[18] for the
massive case (with $m<H_1$), but always in the context of exponential
inflation, which is not the natural inflationary background corresponding
to the low energy string effective action.

The paper is organized as follows.

In Sec.2  we present the general exact solutions (for space-independent
fields)
of the system of
background field equations, including classical string sources, following
from the tree-level string effective action at lowest order in $\ap$. The
explicit form of the solution is displayed, in particular, for a perfect fluid
model of sources, in $D=d+1$ dimensions, for any given equation of state.
The low curvature and large curvature limit of such solutions are given
both in the Brans-Dicke and in the conformally related Einstein frame. In
Sec.3 we derive the coupled system of scalar (metric plus dilaton)
perturbation equations, including the perturbations of the matter sources
in the perfect fluid form. Such equations are applied to compute the scalar
perturbation spectrum for a specific case of background evolution
motivated by a model of sources (presented in Sec.4) in which the dominant
form of matter is a sufficiently diluted, non-interacting gas of large
macroscopic strings. The background describes a phase of growing
curvature and accelerated kinematic (of the pre-big-bang type), which is
expected to evolve towards the standard, radiation-dominated cosmology.
The frame-independence of the inflationary properties of such background
is also discussed in Sec.4. The corresponding spectrum of the produced
dilatons is discussed in Sec.5, where it is shown that, because of its
fast growth with frequency, the phenomenological constraints leave open a
window compatible with the possible dominance of not yet decayed
dilatons (in the hypothesis of negligible classical oscillations of the
dilaton background).
The main results of this paper are finally summarized and
briefly discussed in Sec.6.

\vskip 2 cm

\centerline{\bf 2. General solution of the background field equations.}

We will assume the evolution of the  Universe to be
described at curvatures below the string/Planck scale by the equations
$$
R_\mu\,^\nu +\bigtriangledown_\mu\bigtriangledown^\nu \phi- {1\over
2} \da_\mu\,^\nu {\pa V \over \pa \phi} -{1\over 4} H_{\mu\a
\b}H^{\nu\a\b} = 8\pi G_D e^\phi T_\mu^\nu \eqno(2.1)
$$
$$
R-(\bigtriangledown_\mu\phi)^2+2 \bigtriangledown_\mu
\bigtriangledown^\mu \phi +V-{\pa V\over \pa \phi}-{1\over
12}H_{\mu\nu\a}H^{\mu\nu\a}=0\eqno(2.2)
$$
$$
\pa_\nu(\sqrt{|g|}e^{-\phi}H^{\mu\a\b})=0\eqno(2.3)
$$

Such system of equations follows
 from the low-energy ($D$-dimensional) effective action of
closed (super)string theory [19],
$$
S=-{1\over 16\pi G_D}\int
d^Dx\sqrt{|g|}e^{-\phi}\left[R+\pa_\mu\phi\pa^\mu\phi -{1\over
12}H_{\mu\nu\a}H^{\mu\nu\a}+V(\phi)\right] +S_M\eqno(2.4)
$$
Here $\phi$ is the dilaton field, and $H_{\mu\nu\a}$ the field strength of
the two-index antisymmetric (torsion) tensor $B_{\mu\nu}=-B_{\nu\mu}$.
We have included a possible dilaton potential, $V(\phi)$, and also a
possible phenomenological contribution of the
matter sources represented by
the action $S_M$, whose metric variation produces the stress tensor
$T_{\mu\nu}$.

We shall consider, in this paper, homogeneous backgrounds which are
independent of all spacelike coordinates (Bianchi I type, with $d$ Abelian
isometries), and for which a synchronous
frame exists where $g_{00}=1$, $g_{0i}=0=B_{0i}$ (conventions:
$\mu,\nu=0,1,....D=d+1$; $i,j=1,2,....d$).
 We shall assume, moreover, that
the action $S_M$  describes ``bulk"
string matter, satisfying the classical
string equations of motion in the given background. At tree level
$V$ is a constant. In terms of the ``shifted dilaton"
$$
\fb =\phi -{1\over 2} \ln |det(g_{\mu\nu})|\eqno(2.5)
$$
  the field equations  (2.1)-(2.3) can be
written in matrix form as [20]
$$
\dot {\fb}^2 -2\ddot {\fb} -{1\over 8}\,Tr\, (\dot M\eta)^2 - V=0\eqno(2.6)
$$
$$
\dot {\fb}^2 +{1\over 8}\,Tr\, (\dot M\eta)^2 - V= \rb e^{\fb}\eqno(2.7)
$$
$$
{d\over dt}(e^{-\fb}M\eta \dot M)=\overline T\eqno(2.8)
$$
(a dot denotes differentiation with respect to the cosmic time $t$, and
we have used units in which $8\pi G_D=1$, so that $\rb$ and $\overline T$
have both dimensions $L^{-2}$). Here $M$ is a $2d\times 2d$ matrix,
$$
M= \pmatrix{G^{-1} & - G^{-1}B \cr
BG^{-1} & G-BG^{-1}B \cr}
\eqno(2.9)
$$
where $G$ and $B$ are matrix representation of the $d\times d$ spatial
part of the metric ($g_{ij}$) and of the antisymmetric tensor ($B_{ij}$), in
the basis in which the $O(d,d)$ metric $\eta$ is in off-diagonal form,
$$
\eta=\pmatrix{0 & I \cr I & 0 \cr}\eqno(2.10)
$$
($I$ is the unit $d\times d$ matrix). $\overline T$ is another $2d\times
2d$ matrix representing the spatial part of the string stress tensor [20]
(including the possible contribution of an antisymmetric current density,
source of torsion). Finally $\rb$ is related to the energy density
$\r=T_0\,^0$ by
$$
\rb = \r \sqrt{|det(g_{\mu\nu})|}\eqno(2.11)
$$

The three equations (2.6)-(2.8) correspond, respectively, to to the dilaton
equation (2.2) and to the time and space part of eqs.(2.1), (2.3) for the
homogeneous background that we have considered. Their combination
provides the covariant conservation equation for the source energy
density, which can be written in compact form as [20]
$$
\dot{\rb}+{1\over 4}\,Tr\, (\overline T \eta M\eta \dot
M\eta)=0\eqno(2.12)
$$
The set of equations (2.6)-(2.8), (2.12) is explicitly covariant
 under the
global $O(d,d)$ transformation [21,20]
$$
\fb \ra \fb,\,\,\,\,\,\,\,\,
\rb \ra \rb, \,\,\,\,\,\,\,\,
M \ra \La^T M\La, \,\,\,\,\,\,\, \overline T \ra \La^T \overline T \La
\eqno(2.13)
$$
where $\La$ is an $O(d,d)$ constant matrix satisfying
$$
\La^T \eta \La =\eta \eqno(2.14)
$$
For a suitable class of dilaton potentials such a system can be solved by
quadratures, following the method presented in Ref.[2].

Here we shall concentrate, in particular, on the case $V=0$,
corresponding to strings in critical space-time dimensions (which does
not exclude, however, a description of $d=3$
evolution in case we
add the right number of spectator dimensions in order to compensate
the central charge deficit).
We introduce a suitable
(dimensionless) time coordinate $x$, such that
$$
\rb ={1\over L} {dx\over dt} \eqno(2.15)
$$
($L$ is a constant with dimensions of length), and we define
$$
\Ga = \int ^x {\overline T \over \rb} dx^\pr \eqno(2.16)
$$
The equations (2.6)-(2.8) can then be integrated a first time, with the
help of eq.(2.16) and of the $O(d,d)$ identity
$$
(M\eta \dot M \eta)^2= -(\dot M \eta)^2 , \eqno(2.17)
$$
to give [2]
$$
\rb={e^{\fb} \over 4L^2} D \eqno(2.18)
$$
$$
\fb^\pr =-{2\over D} (x+x_0)\eqno(2.19)
$$
$$
M\eta M^\pr = {4\Ga \over D}\eqno(2.20)
$$
where
$$
D=(x+x_0)^2-{1\over 2}\,Tr\,(\Ga \eta)^2\eqno(2.21)
$$
(a prime denotes differentiation with respect to $x$, and $x_0$ is an
integration constant).

By exploiting the fact that $M$ is a symmetric $O(d,d)$ matrix, $M\eta
M=\eta$, and that, because of the definition of $\overline T$ (see
Ref.[20])
$$
M\eta \Ga =-\Ga \eta M ,\eqno(2.22)
$$
eqs.(2.19), (2.20) can be integrated a second time to give
$$
\fb (x)= \phi_0 -2\int {dx\over D} (x+x_0) \eqno(2.22)
$$
$$
M(x)= P_x \exp (- 4 \int {dx\over D} \Ga \eta ) M_0 \eqno(2.24)
$$
  where $\phi_0$ and $M_0$ are integration
constants ($M_0$ is a symmetric $O(d,d)$ matrix), and $P_x$ denotes $x$-
ordering of the exponential.
For any given ``equation of
state", providing a relation $\overline T =\overline T (\rb)$ between the
spatial part of the stress tensor of the sources and their energy
density, eqs.(2.23) and (2.24), together with (2.18), represent the general
exact solution of the system of string cosmology equations, for
space-independent background fields and vanishing dilaton potential.

Such a solution presents in general singularities, for the curvature and
the effective coupling constant $e^\phi$, occurring in correspondence of
the zero of $D(x)$. It is important to stress that, near the singularity,
the contribution of the matter sources becomes negligible with respect
to the curvature terms in the field equations (just like in general
relativity, in the case of Kasner's anisotropic solution).

The relative importance of the source term is measured indeed by the
ratio (see for instance eq.(2.7))
$$
\Om (x)= -{8\rb e^{\fb} \over (d-1)\,Tr\, (\dot M \eta)^2}\eqno(2.25)
$$
(we have normalized $\Om $ in such a way that it reduces to the usual
expression for the effective energy density in critical units, $\Om =
\r/\r_c$, when the dilaton is constant and the metric isotropic).
According to eqs.(2.18), (2.20)
$$
M\eta \dot M ={e^{\fb}\over L} \Ga \eqno(2.26)
$$
so that, by exploiting the $O(d,d)$ properties of $M$,
$$
Tr\,(\dot M \eta )^2 =-{e^{2\fb} \over L^2} \,Tr\, (\Ga \eta)^2\eqno(2.27)
$$
Therefore
$$
\Om ={2\over d-1} {D\over Tr\, (\Ga \eta)^2} \eqno(2.28)
$$
goes to zero at the singularity ($D\ra 0$).

In this limit, the matter contribution becomes negligible and the general
solution presented here reduces to the $V=0$ case of the general
vacuum solution of the string cosmology equations [22]. Denoting indeed
by $t_c$ a singular point such that $D(t_c)=0$, $\Ga (t_c)\not=0$, from
eq.(2.26) we have, near this point,
$$
M\eta \dot M e^{-\fb} = A\eqno(2.29)
$$
where the constant matrix $A$ satisfies
$$
A={\Ga (t_c)\over L} = -A^T , \,\,\,\,\,\,\,\,\,\,
M\eta  A +A\eta M =0\eqno(2.30)
$$
because of the property (2.22) of $\Ga$. Moreover, from (2.18) and (2.19)
$$
\dot{\fb}^2 = {e^{2\fb}\over 4L^2}(x+x_0)^2\eqno(2.31)
$$
so that, by using (2.27),
$$
\dot{\fb}^2 +{1\over 8}\,Tr\,(\dot M \eta )^2={e^{2\fb}\over 4L^2}D(t_c)
=0\eqno(2.32)
$$
Eqs.(2.29) and (2.32) correspond exactly to the equations defining the
general vacuum solution of Ref.[22], for the case of vanishing dilaton
potential.

Consider now the particular case in which $B_{\mu\nu}=0$, and we are in
a diagonal, but not necessarily isotropic, Bianchi-I type metric
background,
$$
g_{00}=1, \,\,\,\,\,\,\,\,\,\,\,\,\,
g_{ij}=-a^2_i(t)\da_{ij}\eqno(2.33)
$$
(this is the background that will we used here to discuss dilaton
production). The matter sources can be represented in the perfect fluid
form, but with anisotropic pressure,
$$
T_0\,^0= \r,\,\,\,\,\,\,
T_i\,^j= -p_i\da _i\,^j, \,\,\,\,\,\,
p_i/\r = \ga_i = const  \eqno(2.34)
$$
In this case we obtain, from the previous definitions,
$$
M\eta M^\pr = 2 \pmatrix{0 & {a_i^\pr \over a_i} \da_{ij} \cr
-{a_i^\pr \over a_i} \da_{ij}  & 0 \cr}, \,\,\,\,\,
\overline T = \pmatrix{0 & \overline p_i \da _{ij} \cr
-\overline p_i \da _{ij} & 0 \cr}
$$
$$
\ga = \pmatrix {0 & \Ga_i \da_{ij} \cr
-\Ga_i \da_{ij} & 0} ,\,\,\,\,\,\,\,
\Ga_i= \ga_i x +x_i
$$
$$
D= (x+x_0)^2 - \sum_i (\ga_i x +x_i)^2= \a (x-x_+)(x-x_-) \eqno(2.35)
$$
where
$$
\overline p_i = p_i \sqrt{|g|}, \,\,\,\,\,\,\,\,\,\,\,\,\,
\a = 1-\sum_i \ga_i^2
$$
$$
x_{\pm}= {1\over \a} \left\{\sum_i\ga_i x_i -x_0 \pm
\left[(\sum_i\ga_ix_i-x_0)^2+\a(\sum_i x_i^2-x_0^2)\right]^{1/2} \right\}
\eqno(2.36)
$$
and $x_i$, $x_0$ are integration constants. The general solution (2.18),
(2.23), (2.24) becomes explicitly [23]
$$
a_i= a_{0i}|(x-x_+)(x-x_-)|^{\ga_i/\a} |{x-x_+ \over x-x_-}|^{\a_i}
\eqno(2.37)
$$
$$
e^{\fb} = e^{\phi_0}
|(x-x_+)(x-x_-)|^{-1/\a} |{x-x_+ \over x-x_-}|^{-\sigma}
\eqno(2.38)
$$
$$
\rb = {\a \over 4 L^2} e^{\phi_0}
|(x-x_+)(x-x_-)|^{(\a - 1) /\a} |{x-x_+ \over x-x_-}|^{-\sigma}
\eqno(2.39)
 $$
where
$$
\sigma = \sum_i\a_i
\ga_i  , \,\,\,\,
\a_i = {\a x_i+\ga_i(\sum_i \ga_i x_i -x_0)\over
\a \left[(\sum_i\ga_ix_i-x_0)^2+\a(\sum_i x_i^2-x_0^2)\right]^{1/2}}
\eqno(2.40)
$$
and $a_{i0}$, $\phi_0$ are additional integration constants.

This solution has two curvature singularities at $x=x_{\pm}$. Near the
singularity, the presence of matter becomes negligible,
$$
\Om (x) =
{\a (x-x_+)(x-x_-)\over (d-1) \sum_i
(\ga_i x +x_i)^2} \ra 0
\eqno(2.41)
$$
 and one recovers the anisotropic vacuum solution of
string cosmology in critical dimensions [24,25]. Indeed, for $x \ra
x_{\pm}$, one has $|x|\sim |t|^{\a/(1\pm \a \sum \a_i \ga_i)}$, and the
solution behaves like
$$
a_i(t) \sim |t-t_{\pm}|^{\b_i^{\pm}}~,~
\,\,\,\,\,\,\, \fb \sim -\ln |t-t_{\pm}|
\eqno(2.42)
$$
where
$$
\b_i^{\pm}= {x_i \pm \ga_i x_{\pm} \over x_0 + x_{\pm}}
{}~~~~~,~~~~~ \sum_i(\b_i^{\pm})^2 =1 ~.~\eqno(2.43)
$$

In the large $|x|$ (small curvature) limit, on the contrary, the relation
between $x$ and cosmic time is $|x|\simeq |t|^{\a /(2-\a)}$, and the
solution (2.37)-(2.39) behaves like (for $|x|\ra \pm \infty$)
$$
a_i(t) \sim |t| ^{2\ga_i /( 1+\sum \ga_i^2)},\,\,\,\,\,\,
\fb \sim -{2\over 1+\sum \ga_i^2} \ln |t|
$$
$$
\phi \sim 2{\sum \ga_i-1\over 1+\sum \ga_i^2} \ln |t|, \,\,\,\,\,\,
\rb \sim |t|^{-2\sum \ga_i^2/( 1+\sum \ga_i^2)}
\eqno(2.44)
$$
The critical density parameter, in this limit, goes to a constant
$$
\Om_\infty ={1-\sum_i\ga_i^2 \over(d-1)\sum_i\ga_i^2} \eqno(2.45)
$$
which is obviously $\Om_ \infty =1$ for an isotropic, radiation-dominated
background with $\ga_i=1/d$.

It is interesting to point out that, for any solution ${a_i,\phi}$
corresponding to a given set of equations of state, $p_i=\ga_i\r$, there
are the corresponding ``dual" solutions obtained through the reflection
$\ga_i \ra -\ga_i$, which leads to
$
a_i(\ga_i)\ra a_i(-\ga_i)= a_i^{-1}
(\ga_i)
$,
preserving however the values of $\fb$ and $\rb$ (scale-factor
duality [25,26]). Such a duality transformation, combined with the time-
inversion $t \ra -t$, transforms any given metric describing (for $\ga_i
>0$) decelerated expansion with decreasing curvature, $\ddot a_i<0$,
$H_i>0$, $\dot H_i <0$, into a new solution describing (for $\ga_i<0$) a
superinflationary expansion with increasing curvature,
$\ddot a_i>0$,
$H_i>0$, $\dot H_i >0$ (see also [2]).

It is convenient, for later use, to write down explicitly
the isotropic version of the asymptotic backgrounds (2.42) and
(2.44), as a function of the cosmic time $t$ and
conformal time $\eta$ such that $dt=ad\eta$.
In the $(d+1)$-dimensional isotropic case, the
small curvature limit (2.44) becomes, in cosmic time,
$$
a(t)\sim |t|^{2\ga /(1+d\ga^2)},~~~
\phi \sim {d\ga -1\over \ga} \ln a , ~~~
\r \sim a^{-d (\ga +1)} \eqno(2.46)
$$
while in terms of $\eta$ we have
$$
a(\eta)\sim |\eta|^{2\ga/(1-2\ga+d\ga^2)} \eqno(2.47)
$$
The vacuum, dilaton-dominated limit (2.42)
becomes, in the isotropic case,
$$
a_{\mp}(t) \sim |t|^{\mp 1/\sqrt d} , ~~~~~~~
\phi_{\mp} \sim \sqrt d (\sqrt d \pm 1) \ln a \eqno(2.48)
$$
and, in conformal time,
$$
a_{\mp}(\eta)\sim |\eta|^{\mp 1 /(\sqrt d \pm 1)} \eqno(2.49)
$$
Note that for $\ga =1/d$, and $t \ra +\infty$, eq.(2.46) describes the
standard, radiation-dominated cosmology with $\phi=const$; the
dual case, $\ga=-1/d$ and $t \ra 0$, with $t<0$, describes instead a
typical pre-big-bang configuration [2], with a superinflationary
expansion driven by a perfect gas of stretched strings [27]. The dual
solution obtained through a more general $O(d,d)$ transformation,
applied to the radiation case, corresponds to a non-diagonal metric and
an effective viscosity in the source stress tensor, and has been
discussed in Ref.[20].

We note, finally, that the solution presented in this section is given
explicitly in the Brans-Dicke (BD) frame, whose metric coincides with
the $\sg$-model metric to which strings are directly coupled. The passage
to the Einstein (E) frame, defined as the frame in which the graviton
and dilaton kinetic terms are diagonalized, and the action takes the
canonical form
$$
S_E={1\over 16 \pi G_D} \int d^Dx \sqrt{|g_E|} \left[-R(g_{\mu\nu}^E) +
{1\over 2} g_E^{\mu\nu} \pa_\mu \phi_E \pa_\nu \phi_E\right] \eqno(2.50)
$$
is obtained through the conformal rescaling
$$
g^E_{\mu\nu} = g_{\mu\nu} e^{-2\phi/(d-1)},~~~~~~
\phi_E =\sqrt{{2\over d-1}} \phi \eqno(2.51)
$$
The E-transformed scale-factor, $a_E$, and the cosmic time coordinate,
$t_E$, are thus related to the original BD ones by
$$
a_E= a e^{-\phi/(d-1)}, ~~~~~~~~~
dt_E= dt e^{-\phi/(d-1)} \eqno(2.52)
$$

The asymptotic limit (2.46) of the previous general solution thus becomes,
in the E frame,
$$
a_E(t_E) \sim |t_E|^\beta,~~~~~~~~~~
\phi_E \sim \sqrt{{2\over d-1}} {(d-1)(1-d\ga)\over \ga-1} \ln a_E
$$
$$
\r_E \sim a_E^{-2/\b},~~~~~~~~~~~~~
\b={2(1-\ga)\over (d-1)(1+d\ga^2)-2(d\ga-1)}\eqno(2.53)
$$
where $\r_E$ is conformally related to the BD energy density $\r$ by
$$
\r_E =\r {\sqrt{|g|}\over \sqrt{|g_E|}} = \r e^{\phi(d+1)/(d-1)}
\eqno(2.54)
$$
In conformal time,
$$
a_E(\eta)\sim |\eta|^{-2(\ga-1)/(d-1)(1-2\ga+d\ga^2)} \eqno(2.55)
$$
(note that the conformal time coordinate is the same in the E and BD
frame,
$$
d\eta_E={dt_E\over a_E(t_E)}={dt\over a(t)}= d\eta \eqno(2.56)
$$
because of eq.(2.52)). The high curvature limit (2.48) becomes, in the E
frame,
$$
a_{\mp}^E(t_E) \sim |t_E|^{1/d}, ~~~~~~~~~~~
\phi_{\mp}^E \sim \mp \sqrt{2d(d-1)} \ln a_E \eqno(2.57)
$$
and, in conformal time,
$$
a_{\mp}^E(\eta_E) \sim |\eta|^{1/(d-1)}  \eqno(2.58)
$$

It should be stressed that the radiation-dominated solution, with $\ga=1
/d$, $\phi=const$, is obviously the same in both frames, see eqs.(2.53)
and (2.48). We note also that in the vacuum, dilaton-dominated case, the
duality transformation which is represented in BD frame by the inversion
of the scale factor, with related dilaton shift,
$$
a_\pm \ra a_\mp =a_\pm^{-1},~~~~~~~~
\phi_\pm \ra \phi_\mp = \phi_\pm - 2d \ln a _\pm \eqno(2.59)
$$
in the E frame becomes a transformation between the weak coupling and
the strong coupling regime,
$$
\phi_\pm^E \ra \phi_\mp^E= - \phi_\pm^E, \eqno(2.60)
$$
without changing the metric background.

Concluding this section, we want to stress that the solutions discussed
so far describe the situation in which the dilaton potential can be
neglected, namely the background evolution at early enough times when
the effective coupling $e^\phi$ is small enough. Indeed, because of non-
renormalization theorems, the potential is expected to appear at the
non-perturbative level only, and has to be extremely small ($V(\phi)
\sim exp[-exp(\phi)]$) in the weak coupling regime. At later times, and
large couplings, the main effect of the dilaton potential
will be taken
into account in the form of a dilaton mass term (see Sections 3,5),
which freezes the Newton constant at its present value.

\vskip 2 cm

\centerline{\bf 3.
Scalar perturbations with dilaton and perfect fluid sources.}

In order to obtain the equations governing the classical evolution of
  scalar perturbations, we choose to work in the Einstein frame, where
the explicit form of the equations is simpler.
 This   is a legitimate choice
 since, as we shall see at the end of this
Section, the scalar perturbation spectrum, just like the
 graviton spectrum [23], is the same in the Einstein and  Brans-Dicke
frames.

In the E frame, the background field equations (with $B_{\mu\nu}=0$ but
with a non-vanishing dilaton potential $V$) take the form
$$
2R_\mu~^\nu-\da_\mu^\nu R = \pa_\mu \phi \pa^\nu \phi +\da_\mu^\nu
\left[V-{1\over 2}(\big \phi)^2\right] +T_\mu~^\nu \eqno(3.1)
$$
$$
\big_\mu \big^\mu \phi + {\pa V\over \pa \phi} + c T =0 \eqno(3.2)
$$
where $c=\sqrt{2/(d-1)}$ (the Einstein frame index, E, will be omitted
throughout this Section). The coupling of the dilaton to the stress
tensor of the matter sources is fixed by the conformal rescaling (2.51),
(2.54).
We start, for simplicity, with a
($d+1$)-dimensional isotropic background, with perfect fluid sources,
$$
g_{\mu\nu} = diag (1, -a^2 \da_{ij}),~~~~~~~~~~~~
\phi= \phi(t)
$$
$$
T_\mu~^\nu =(\r+p)u_\mu u^\nu -p \da_\mu^\nu,~~~~~~~~~~~~~~
u^\mu =\da_0^\mu
\eqno(3.3)
$$
and we consider the pure scalar part of the metric perturbations, $\da
g_{\mu\nu} \equiv h_{\mu\nu}$, together with the perturbations of the
dilaton background, $\da \phi \equiv \c$, and of the matter sources,
$\da \r , \da p, \da u^\mu$ (in the linear approximation scalar, vector
and tensor perturbations are decoupled, and evolve independently). We
use here for the metric  the Bardeen variables $\Phi$, $\psi$, that are
invariant under those infinitesimal coordinate transformations which preserve
the scalar nature of the fluctuations [28-30]. In the longitudinal gauge
we have thus the first-order expressions [30]
$$
h_{00} = 2\Phi = h^{00},~~~~~~~~~~~~~~~~~~~~
h_{0i}=0
$$
$$
h_{ij}=2a^2\psi \da_{ij},~~~~~~~~~~~~~~~~~~~~
h^{ij}={2\over a^2}\psi \da^{ij}
$$
$$
\da T_0~^0=\da \r,~~~~~~~~~~~~~
\da T_i~^j =-\da_i^j \da p , ~~~~~~~~~~~~~
\da T_i~^0 ={\r +p \over a} \da u_i \eqno(3.4)
$$
These are to be inserted into the first order perturbation of the
Einstein equations (3.1),
$$
-2 h^{\nu\a} R_{\mu\a} +2g^{\nu\a} \da R_{\mu\a}- \da _\mu^\nu
(g^{\a\b}\da R_{\a\b}-h^{\a\b}R_{\a\b})=
$$
$$
=-h^{\nu\a}\pa_\mu\phi \pa_\a\phi + \pa_\mu\c \pa^\nu\phi
+ \pa_\mu\phi \pa^\nu\c +\da_\mu^\nu {\pa V\over \pa \phi}\c-
$$
$$
-{1\over 2}
\da_\mu^\nu (2g^{\a\b}\pa_\a\phi \pa_\b\c -h^{\a\b}\pa_\a\phi\pa_\b\phi)
+\da T_\mu~^\nu \eqno(3.5)
$$
and of the dilaton equation (3.2)
$$
-h^{\a\b}\big_\a\big_\b \phi +\big_\mu\big^\mu \c
-g^{\a\b}(\da \Ga_{\a\b}~^\mu)\pa_\mu \phi +{\pa^2V\over \pa\phi^2} \c
+{c\over 2} (\da \r -d\da p)=0\eqno(3.6)
$$
Here the covariant derivatives are to be performed with respect to the
background metric $g_{\mu\nu}$, and $\da R_{\mu\nu},\da \Ga$ are to be
computed to first order in $h_{\mu\nu}$.

By using the background field equations, the ($i,j$) component of
eq.(3.5), with $i\not=j$, gives
$$
\Phi=(d-2)\psi \eqno(3.7)
$$
which allows us to eliminate everywhere one of the two Bardeen's
variables. The ($i,0$) component
gives a constraint which can be written, in terms of the conformal time
$\eta$, as
$$
\pa_i\left\{2(d-1)\left[\H(d-2)\psi +\psi^\pr\right]-\c \fp \right\}
=(\r+p)a\da u_i \eqno(3.8)
$$
(a prime denotes differentiation with respect to $\eta$). The ($0,0$)
component provides an expression for the density perturbation $\da \r$
in terms of the scalar variables $\psi$ and $\c$,
$$
\big^2\psi -d\H\psi^\pr -\left[d(d-2)(\H)^2 -{d-2\over 2(d-1)} \fp^2
\right]\psi=
$$
$$
={1\over 2(d-1)}(\fp \c^\pr +{\pa V\over \pa \phi} a^2\c + a^2\da \r )
\eqno(3.9)
$$
Finally, the ($i,i$) component of eq.(3.5) and the perturbed dilaton
equation (3.6) give, respectively,
$$
\psi^\se +(2d-3)\H\psi^\pr+\left\{(d-2)\left[2{a^\se
\over a} +(d-4)(\H)^2\right]+
{d-2\over 2(d-1)}\fp^2 \right\} \psi=
$$
$$
={1\over 2(d-1)}(\fp \c^\pr -{\pa V\over \pa \phi} a^2\c + a^2\da p)
\eqno(3.10)
$$
$$
\c^\se +(d-1)\H\c^\pr + (a^2{\pa^2 V\over \pa\phi^2}-\big^2)\c=
$$
$$
=2(d-1)\fp \psi^\pr -2(d-2)\left[{c\over 2}(\r - d p) +
{\pa V\over \pa \phi}\right]
a^2\psi -{c\over 2}a^2(\da \r -d \da p)\eqno(3.11)
$$

The linear system formed by the four coupled equations (3.8)-(3.11)
determines the classical evolution of four independent perturbation
variables $\psi,\chi,\da \r$ and $\da u$ (an additional relation between
$\da p$ and $\da \r$ is to be provided by the detailed model of matter
sources). In the absence of dilaton background ($\phi=0=\c$) one
recovers the usual system of equations for hydrodynamical perturbations
[30], while in the absence of fluid sources ($T_\mu^\nu=0=\da
T_\mu^\nu$)
one has the usual perturbation system for a scalar field minimally
coupled to the metric [30].

When $\phi$ and $T_{\mu\nu}$ are both non-vanishing, and $\da p$ can be
parametrized in terms of $\da \r$ as $\da p= \ep(t)\da \r$, we may
eliminate $\da \r$ by means of eq.(3.9), and the system
 reduces to a pair of
 second-order differential equations (3.10), (3.11) for the coupled
variables $\psi$ and $\chi$. By introducing the bi-dimensional vector
$$
Z=\pmatrix {\psi \cr \c \cr}\eqno(3.12)
$$
and by  the dilaton background as
$$
\phi = \b \ln a,~~~~~~~~~~~~~~~~~~~~~~ \b= const\eqno(3.13)
$$
the abovementioned pair of equations can be represented in compact form as
$$
Z^\se_k +2\H A Z_k^\pr +(k^2B+C)Z_k=0\eqno(3.14)
$$
where $Z_k=(\psi_k,\c_k)$ represents the Fourier component of the
perturbation variables, $\big^2 Z_k= -k^2 Z_k$, and
$$
A=\pmatrix{{1\over 2}(2d-3+d\ep) & -{1-\ep \over 4(d-1)}\b \cr
-(d-1)[{cd\over 2}(1-d\ep)+\b] & {1\over 2}(d-1)-{c\over 4}(1-d\ep)\b
\cr}
$$
$$
B=\pmatrix{\ep & 0\cr -c(1-d\ep)(d-1) & 1 \cr}
$$
$$
C=\pmatrix{2(d-2){a^\se \over a} +(d-2)[d-4+d\ep + {1-\ep \over 2(d-1)}
\b^2](\H)^2 & 0 \cr
-cd(d-2)(\ga-\ep)[d(d-1)-{\b^2\over 2}](\H)^2 & 0 \cr}
\eqno(3.15)
$$
(we have neglected here the possible contribution of the dilaton
potential, by putting $\pa V/\pa\phi=0=\pa^2 V/\pa \phi^2$).
We note that
a system of coupled scalar perturbation equations similar to (3.14) was
previously considered also in Refs.[31,32] where, however, a scalar
field model of source (``inflaton" matter) was used, instead of the
fluid model adopted in this paper.

Without further approximations, $\psi$ and $\c$ are thus in general
non-trivially mixed, with time-dependent mixing coefficients
determined by the explicit model of sources, $\ga= p/\r$, $\ep= \da p/
\da \r$, and by the background kinematics, $a(t)$, $\phi(t)$, according
to eq.(3.14). The solution of (3.14) provides in turn, for any given
background configuration, a unique determination of the density contrast
$\da \r /\r$ through eq.(3.9), and of the velocity perturbation $\da
u_i$ through eq.(3.8).

Eqs.(3.8)-(3.11) are linear in the perturbations and just describe
their classical evolution without specifying their absolute magnitude.
 As clearly stressed in Ref.[30]
 (see also Refs.[33,34]), in order
to determine the absolute magnitude
  of the vacuum fluctuations and their
spectral distribution, one must express the perturbations in terms of
the correctly normalized variables satisfying canonical commutation
relations. These can be determined by expanding the action to
second order in the fluctuations.

For
the pure metric-scalar field system ($T_\mu^\nu=0$) such a canonical
variable is known to be fixed by the following
linear combination of $\c$ and
$\psi$ [35-37]
$$
v=a\c +z\psi,~~~~~~~~~~~~~~~~~~~
z={a^2\fp \over a^\pr}. \eqno(3.16)
$$
For a pure fluid source ($\phi=0$), with constant $\ep$, the canonical
variable is instead [38,39]
$$
w={1\over 6}(f-\xi \psi),~~~~~~~~~~~~~~~~~~~
\xi={a^3\over a^\pr} \sqrt{{\r +p\over \ep}}\eqno(3.17)
$$
where $f$ is the velocity potential determining the fluid perturbations
as
$$
\eqalign{&
a\sqrt{\r +p}~\da u_i = -\sqrt{\ep}~ \pa_i f \cr
& a^2 \da \r =-{a^2 \over \ep}(\r +p) \psi -{1\over a^2}\left({a^2 f
\sqrt{\r + p} \over \sqrt \ep}\right)^\pr \cr} \eqno(3.18)
$$
(we have assumed $d=3$ in the previous three equations). The variables
$v,w$ play the role of ``normal coordinates" decoupling the system of
perturbation equations, and reducing the action to the free scalar field
form [30]. Only when $\psi$ is expressed in terms of such variables one
gets a canonical normalization of the Fourier modes $\psi_k$, and then
the correlation function for the metric fluctuations
$$
\langle \psi(x)\psi(x^\pr) \rangle = \int
{dk\over k}{\sin kr \over kr} |\da_\psi (k)|^2 \eqno(3.19)
$$
provides the correct spectral distribution for the metric
$$
|\da_\psi(k)|^2 = k^3|\psi_k|^2 \eqno(3.20)
$$
and for the dilaton, $\da_\c (k)$, through eq.(3.16).

If $T_\mu^\nu$ and $\phi$ are both non-vanishing, one could try a
perturbative approach to the spectrum (like in Refs.[32,40]), by keeping
fixed the definitions of $v$ and $w$ as a zeroth-order approximation. In
such case, the constraint (3.8) gives (in $d=3$)
$$
\psi^\pr +\H \psi = {1\over 4}\fp \c -{1\over 4}\sqrt{\ep (\r +p)} f
\eqno(3.21)
$$
By eliminating $f$ and $\c$ in terms of $v$ and $w$ through eqs.(3.16),
(3.17), by using the constraint (3.21) and the background field
equations, one can then express the Fourier mode $\psi_k$, from
eq.(3.9), as
$$
\psi_k=\psi_k(v,v^\pr,w,w^\pr,k) \eqno(3.22)
$$
Moreover, the system of equations formed by eq.(3.11) and by the
combination of eqs.(3.9), (3.10) obtained by eliminating $\da \r$, can be
written as a system of two second-order differential equations for the
coupled modes $v_k$ and $w_k$. Its solutions, when inserted into
eq.(3.22), provide a first approximation to the scalar perturbation
spectrum (3.20). From eq.(3.17) one has then the corresponding dilaton
spectrum, $|\da_\c|=k^{3/2}|\c_k|$, and from eqs.(3.18), (3.17) the
density perturbation spectrum $|\da_\r|=k^{3/2}|(\da \r /\r)_k|$.

In general, dilaton and metric fluctuations will have different spectral
distributions, $|\da_\psi| \not= |\da_\chi|$. The coupled system of
equations is rather complicated, but it seems possible, in principle, to
obtain a large variety of spectra as the equation of state and the ratio
$\da p/\da \r$ are appropriately varied [41].

In this paper we shall consider a model (see Sec.4 for its motivations) in
which the universe evolves from a three-dimensional, dilaton-dominated
phase of the pre-big-bang type (with negligible fluid sources, $T_\mu^\nu=0=
\da T_\mu^\nu$), to the standard radiation-dominated phase ($p=\r /3$),
adiabatic ($\ep= 1/3$), and with frozen Newton constant ($\phi= const$).
More complicated scenarios will be analyzed in future works [41]. The
phase of pre-big-bang inflation is assumed to extend in time from $-
\infty$ up to the time $\eta=-\eta_1 <0$, which marks a sudden
transition to the phase of radiation dominance. For $\eta< -\eta_1$ the
constraint (3.8) thus becomes
$$
\psi_k^\pr + \H \psi_k^\pr = {1\over 4}\fp \c_k \eqno(3.23)
$$
where, according to eq.(3.16),
$$
\c_k ={v_k\over a} -{z\over a}\psi_k \eqno(3.24)
$$
When the constraint is inserted into eq.(3.9), and eq.(3.24) is used in
order to eliminate $\c$ and $\c^\pr$, we are led to a relation of the
form (3.22), namely
$$
\psi_k= -{1\over 4 k^2} \fp {z\over a}\left(v_k\over z\right)^\pr
\eqno(3.25)
$$
In the absence of matter sources, eq.(3.11) becomes equivalent to the
combination of eqs.(3.9) and (3.10). By expressing $\c_k$ in terms of
$v_k$ according to eq.(3.24), and by eliminating $\psi^\pr, \psi^\se$
through eq.(3.23), we finally get the canonical perturbation equation
[30], valid for $\eta<-\eta_1$,
$$
v_k^\se +(k^2-{a^\se \over a})v_k =0 \eqno(3.25)
$$

In the second, radiation-dominated phase
($\eta >-\eta_1$), we assume that
the dilaton acquires a mass $m$, and it stays frozen at the minimum of
the potential (with possible small oscillations around it), so that
$$
V=0={\pa V\over \pa \phi},~~~~~~~~~~~~~~~~~~~~~~~
{\pa^2 V\over \pa \phi^2}= m^2 \eqno(3.26)
$$
In this case $\chi$ decouples from the metric fluctuations (see
eqs.(3.8)-(3.11)), that are coupled now to the fluid perturbations only;
the canonical variable for their quantization is thus given by eq.(3.17).
As $a^\se/a =0$ in the radiation phase, it turns out however that for
$\eta >-\eta_1$ both $w$ and $a\c$ satisfy the free oscillator equation,
$w^\se /w = const$ (apart from the dilaton mass term, assumed to be
negligible at early enough times, see Sec.5). As a consequence, $\psi$
and $\c$ will have the same spectrum (identical, in this case, to the
tensor perturbation spectrum), which can be computed by adopting a
second quantization approach, regarding the amplification of the
perturbations as a process of particle production from the vacuum, under
the action of the cosmological background fields [30].

The Bogoliubov coefficients $c_\pm$ for such a process are obtained by
matching the solution of eq.(3.25) to a
general solution of the plane-wave type,
$$
v_k={1\over \sqrt{k}}(c_+e^{-ik\eta}+c_-e^{ik\eta}) \eqno(3.27)
$$
valid for $\eta>-\eta_1$. By assuming, for $\eta<-\eta_1$, that
$$
a \sim (-\eta)^{-\a},~~~~~~~~~~~~~~~~~~~~~~~~~~~
\phi =\b \ln a \eqno(3.28)
$$
we have
$$
{z^\se \over z}= {a^\se \over a}= {\a(\a+1)\over \eta^2} \eqno(3.29)
$$
The solution of eq.(3.25) describing oscillations with positive
frequency at $\eta=-\infty$, and defining the initial vacuum state, is
thus given in terms of the second kind Hankel function, $H^{(2)}$, as
$$
v_k= \eta^{1/2} H_\nu^{(2)}(k\eta),~~~~~~~~~~~~~~~~~~~~~~
\nu = \a+{1\over 2} \eqno(3.30)
$$
The continuity of $v_k$ and $v_k^\pr$ at the transition time $\eta=
-\eta_1$ fixes the Bogoliubov coefficient $c_-(k)$, and the
corresponding expectation number of particles produced in the mode $k$.
For $k \me 1$ we obtain
$$
\langle n(k) \rangle= |c_-(k)|^2 \simeq (k\eta_1)^{-2|\nu|-1}
\eqno(3.31)
$$
(higher mode production turns out to be exponentially suppressed [3,
42], and can be neglected for the purpose of this paper).
In terms of the proper frequency $\om = k/a$, the energy density
$\r_\c$ of the produced dilatons is thus characterized by a spectral
distribution
$\om d\r_\c /d\om\simeq \om^4 |c_-|^2$ which may be written, in units
of critical density $\r_c = H^2/G$,
$$
\Om_\c (\om, t) ={\om \over \r_c}{d\r_\c \over d\om} \simeq
{G \om^4 \over H^2} \left (\om \over \om_1\right)^{-2-2\a} =
GH_1^2\left(\om\over \om_1\right)^{2-2\a}\left(H_1\over H\right)^2
\left(a_1\over a\right)^4
\eqno(3.32)
$$
where $\om_1(t)=H_1a_1/a(t)$ is the proper frequency of the highest
excited mode (here we have supposed $\a \geq -1/2$). This is the same
spectrum as that obtained in the graviton case [43], with an intensity
normalized to the final inflation scale $H_1$. It is growing for a phase
of superinflationary pre-big-bang expansion ($\a<1$), flat for de Sitter
($\a=1$), and decreasing for power-law inflation ($\a>1$).

It should be stressed that this second quantization approach
is convenient to discuss the squeezing properties of the
produced radiation [3,34,44-47] but,
for what concerns the perturbation spectrum,
it is completely equivalent to the
more traditional approach in which one computes the parametric
amplification of the perturbation amplitude. In this second approach one
has indeed,
according to the
``effective potential" $z^\se/z$ of eq.(3.25), a mode
amplitude which is constant, $|v_k|\simeq 1/\sqrt k$, in the initial
region $\eta \ra -\infty$ where $k^2>>|z^\se /z|\simeq \eta^{-2}$, and
which grows with power-like behaviour in $\eta$ in the non-oscillatory
region defined by $k^2<<|z^\se /z|$ (in the subsequent radiation era the
solution for $v$ is again oscillating, with frozen amplitude). In the
non-oscillatory region
$$
v_k= c_1 z + c_2 z \int^\eta {d\eta^\pr \over z^2(\eta^\pr)} -
k^2  z \int^\eta {d\eta^\pr \over z^2(\eta^\pr)}
\int^{\eta^\pr} {d\eta^\se  z^2(\eta^\se)}
+ O(k^4) \eqno(3.33)
$$
($c_1, c_2$ are integration constants) is the general solution of
eq.(3.25) to first order in $k^2$ (the first sub-leading term has been
included to have non-trivial derivative of $v/z$). This gives, for the
background (3.28) (with obvious redefinition of $c_1,c_2$, and
introducing a further numerical constant $c_3$),
$$
v_k=c_1|\eta|^{-\a} + c_2 |\eta|^{1+\a}-c_3k^2|\eta|^{2-\a} \eqno(3.34)
$$
For $\a >0$ (inflationary expansion) the first term is the dominant one
in the the $|\eta|\ra 0$ limit, and
the wave amplification achieved in
this limit can thus be estimated as [30]
$$
v_k \simeq {v_k(\eta)\over \sqrt k}
\left(1\over v_k\right)_{k\simeq aH}
\simeq {z\over \sqrt k}
\left(1\over v_k\right)_{k\simeq aH}
[1-{k^2 \eta^2\over 2(1-2\a)}]
\eqno(3.35)
$$

The variable at the denominator is to be evaluated at the time $\eta
\simeq k^{-1}$, where the mode $k$ ``hits" the effective potential
barrier $z^\se /z\simeq \eta^{-2}$ (otherwise stated: at the time of
first horizon crossing, when $H=\om$). By inserting this value of $v$
into eqs.(3.25) and (3.20), and recalling the definition of $z$, we are
led to
$$
|\da_\psi(k)|^2 \simeq k^3 |\psi_k|^2 \simeq \left(k\over z
\right)^2_{k\simeq aH}
\simeq \left(H^2\over \dot \phi \right)^2_{k\simeq aH} \eqno(3.36)
$$
which is the standard expression for the scalar perturbation spectrum
[35,48] associated with the inflation-radiation transition (see Sec.4 for
a proof of the fact that the same result is recovered in the
case of contracting backgrounds). The same spectrum is obtained for the
dilaton perturbations as we have, from eq.(3.24),
$$
|\da_\c|^2=k^3|{v_k\over a}-{z\over z}\psi_k|^2
\simeq \left(k\over z\right)^2_{k\simeq aH} \simeq |\da_\psi|^2 \eqno(3.37)
$$

It is important to stress that this expression,
 when multiplied by $G$,  exactly
coincides with the spectral energy density (3.32) (modulo
numerical factors of order unity), evaluated in the radiation era. Indeed,
 multiplying and dividing eq.(3.37) by $H_1^2 \simeq (a_1\eta_1)^{-2}$ we
have
$$
G|\da_\c|^2 \simeq G\left(H^2\over \dot \phi\right)^2_{k\simeq aH}
\simeq \left(G\over a\eta\right)^2_{k\simeq aH} \simeq GH_1^2
\left(a_1\eta_1\over a\eta\right)^2_{k\simeq aH}
$$
$$
 \simeq GH_1^2 (k\eta_1)^{2-2\a}
= GH_1^2\left(\om\over \om_1\right)^{2-2\a} \simeq
\left({\om \over \r_c}{d\r_\c \over d \om}\right)_{rad} \eqno(3.38)
$$
in agreement with eq.(3.32) for $a\sim t^{1/2}\sim H^{-1/2}$.

As already stressed in Ref.[14] for the tensor perturbation case, we want
to remark finally that the scalar perturbation spectrum is the same in
the E and BD frame, as a consequence of the equality of the two
conformal time coordinates (see Sec.2). Indeed, quite independently of the
computational method (first or second quantization) the spectral
behaviour of the energy density is fixed by the Bessel index
$\nu$ of the solution of eq.(3.25), which depends, in  turn, on the
slope of the effective potential $z^\se/z$. For a generic  $d=3$
background in the E frame we have (recall eq.(2.55))
$$
\a(\ga)={\ga -1\over 1-2\ga +3\ga^2}\eqno(3.39)
$$
so that, according to eq.(3.29)
$$
\left(z^\se\over z\right)_E=\left(a^\se\over a\right)_E=
{\ga\over \eta_E^2}{(\ga-1)(3\ga-1)\over (1-2\ga+3\ga^2)^2}\eqno(3.40)
$$
Eq.(3.25) for $v_k$ is not conformally invariant, and in the BD frame
the effective potential becomes
$$
\left(z^\se\over z\right)_{BD} =\left(z^\se\over z\right)_E(a_{BD},
\phi_{BD}) =
\left(a^\se\over a\right)_{BD} -\fp_{BD}\left(\H\right)_{BD} -{1\over 2}
\phi_{BD}^\se +{1\over 4}\fp_{BD}^2 \eqno(3.41)
$$
In this frame, however, the spectrum is determined by the conformally
transformed backgrounds, namely by the solutions (2.46),(2.47) of the BD
field equations. By inserting their explicit expressions for $d=3$ we
get
$$
\left(z^\se\over z\right)_{BD}=
{\ga\over \eta_{BD}^2}{(\ga-1)(3\ga-1)\over (1-2\ga+3\ga^2)^2}\eqno(3.42)
$$
which coincides with the effective E-frame potential (3.40) because of
the equality $\eta_{BD}=\eta_E$. The same results holds for a
dilaton-driven evolution, described by the solution (2.57) and their
BD-transformed expressions.

\vskip 2 cm

\centerline{\bf 4.
Pre-big-bang scenario in the Brans-Dicke and Einstein frame.}

As seen in the previous Section, the spectral distribution of the
perturbations is uniquely fixed by the explicit form of the background
solution. The time-evolution of the background fields   is
determined, in   turn, by the particular model of matter sources. As in
our previous work [20,27], our model of
sources consists of a sufficiently diluted gas of
 classical fundamental strings
whose mutual interactions are described, in a mean-field
approximation sense, as the interaction of each single string
 with the background
generated by all the others  according to the tree-level
effective action (2.4). The source stress-tensor appearing
 in eq.(2.1) is thus
given by a sum over all strings (labelled by $i$) of the
 stress-tensor of each individual string $T_i^{\mu\nu}$, where
$$
T_i^{\mu\nu}(x)= {1\over \pi \ap \sqrt{|g|}}\int d\sg d\tau (
{dX_i^\mu \over d\tau} {dX_i^\nu \over d\tau}-
{dX_i^\mu \over d\sg} {dX_i^\nu \over d\sg}) \da^D (X-x) \eqno(4.1)
$$
and, for each $i$, the  coordinates $X^\mu$  satisfy the   string
 equations of motion in the given background,
$$
{d^2X^\mu \over d\tau^2} - {d^2X^\mu \over d\sg^2} +\Ga_{\a\b}^\mu
({dX^\a \over d\tau} + {dX^\a \over d\sg} )({dX^\b \over d\tau} -
{dX^\b \over d\sg} )=0
$$
$$
g_{\mu\nu}({dX^\mu \over d\tau} {dX^\nu \over d\tau} +
{dX^\mu \over d\sg} {dX^\nu \over d\sg} )=0,\,\,\,\,\,
g_{\mu\nu}{dX^\mu \over d\tau} {dX^\nu \over d\sg} =0 \eqno(4.2)
$$
Here $(2 \pi \ap)^{-1}$ is the string tension, $\Ga_{\mu\nu}^\a$ the
Christoffel
symbol for the background metric $g_{\mu\nu}$, $\tau$ and $\sg$ the usual
world-sheet time and space variables (we are using the gauge in which the
world-sheet metric is conformally flat).

The general exact solution of the system of equations
 (2.1)-(2.3), (4.1), (4.2)
is hard to find and certainly impossible to express in closed form.
 In some
appropriate asymptotic regime, however, the solution of the string
equations of motion, when inserted into the energy-momentum tensor (4.1),
provides an effective equation of state which allows
 to describe
the string sources in the perfect fluid approximation [20,27],
 and to recover
then the general background solutions of Sec.2. The cosmological solution
we are looking for is characterized in particular by having, as initial
configuration, the string perturbative vacuum, namely flat spacetime with
vanishing torsion and coupling constant, $H_{\mu\nu\a}=0$, $\phi=-\infty$.
In this regime strings  move freely, do not decay and behave as
a pressureless  gas
with an energy density $\r$. We shall assume $\r$ to be small enough
initially so that, as we shall see, it will represent a
negligible source of curvature. On the other hand a finite $\r$ is
certainly sufficient to make the dilaton evolve away from the
perturbative minimum.

Indeed, the negative branch ($x\leq x_-$) of the general background solution
 with
perfect fluid sources, eqs.(2.37)-(2.31),
 may be written in the case of vanishing pressure ($\ga_i=0$) as
$$a_i(t)=a_{i0}|{t-2T\over t}|^{\a_i},\,\,\,\,\,\,
e^{\fb} ={16L^2e^{-\phi_0}\over |(t(t-2T)|},\,\,\,\,\,\,
\rb ={1\over L}{dx\over dt}= {e^{\phi_0} \over 4L^2}= const
$$
$$
\a_i={t_i\over T},\,\,\,\,\,\,\,\,
T=\left(\sum_it_i^2\right)^{1/2},\,\,\,\,\,\,\,\, t\leq 0 \eqno(4.3)
$$
($t_i$ are integration constants, and we have performed a time translation
to shift the singularity from $x=x_-$ to the origin, by choosing
$x_0=-T(e^{\phi_0}/4L)$). This background is certainly consistent with the
solution of the string equations of motion (4.2) in the $t \ra  -\infty$ limit.
Indeed, in this limit, the metric is flat,
$$
a_i= const,\,\,\,\,\,\,\, \phi \sim -2\ln(-t), \,\,\,\,\,\,\, \r= const
 \eqno(4.4)
$$
and the solutions of eqs.(4.2) are characterized by $\sum_i(dx^{i}/d\tau)^2=
\sum_i(dx^{i}/d\sg)^2$. Eq.(4.1) gives then $T_0^0=const$, $T_i^{i}=0$,
namely a stress tensor describing dust-like matter in the perfect fluid
approximation. For $t\sim -T$, however, the curvature scale begins to
increase, the string sources progressively enter a non-oscillating unstable
regime [27], and one must then take into account the fact that the
ratios $\ga_i=p_i/\r$ begin to evolve in time.

In connection with this last  point we note that the solution (4.3),
which,
for $t \leq 0$, gives
$$
H_i={2t_i \over t(t-2T)}, \,\, \,\,\,\dot {\fb} =-{2(t-T)\over
 t(t-2T)},\,\,\,\,\,
\rb e^{\fb}={4\over t(t-2T)}= \r e^\phi \; , \eqno(4.5)
$$
is characterized by two scales. One is the curvature scale ($|H_T|\sim
T^{-1}$ at $t\sim -T$) at which the transition from flat to curved
spacetime regime occurs and inflation begins.
 At $t\sim -T$, the curvature $H^2$ is of the same order
as  $\dot {\fb}^2$ or $\r e^\phi$ while, much earlier, it was
negligible. By contrast,
much later than $ t = - T$, it is $\r e^\phi$ which becomes negligible and
one recovers the vacuum solutions. $T$
  is a free phenomenological parameter of the
solution. The other scale is the  maximal scale $|H_1|\sim t_1^{-1}$, at the
time $t\sim -t_1$, after which the solution is no longer valid, because higher
orders in $\ap$ have to be added to the low energy
effective action (2.4). This final scale $t_1$ is thus determined by the string
tension as $t_1\simeq \sqrt {\ap} = \la_s$, where $\la_s$ is the fundamental
(minimal) length parameter of string theory [49], which may be
assumed to coincide  roughly with the present value of the Planck length,
 $\l_p
= M_p^{-1}$. The important point to stress is that,
 in any realistic inflationary scenario,
 $T$ and $\la_s$ cannot be of the same order as we will now show.

When $|t|<T$, the solution describes an accelerated evolution given
asymptotically by
$$
a_i(t) \sim (-t)^{-\a_i},\,\,\,\,\,\,\, |\a_i|<1,\,\,\,\,\,\,\, \sum
_i\a_i^2=1\eqno(4.6)
$$
and which is of the type given in eq.(2.42) (we call ``
accelerated" a configuration in which
$\dot a$, $\ddot a$ and $\dot H$ have the same sign, positive for
expansion, negative for contraction [2,23,27]). In this metric,
the particle horizon along any given spatial direction,
$$
d_p^{i}(t)= a_i(t)\int _{-\infty}^t dt^\pr a_i^{-1}(t^\pr) \eqno(4.7)
$$
evolves for $|t|<<T$ like the scale factor, $d_p^{i}\sim a_i$, while the event
horizon
$$
d_e^{i}(t)= a_i(t)\int _{t}^0 dt^\pr a_i^{-1}(t^\pr) \eqno(4.8)
$$
shrinks linearly in time, $d_e^{i}\sim (-t)$, for $t\ra 0$. The ratio of the
two
proper sizes $r^{i}(t)=d_p^{i}/d_e^{i}$ thus grows in time, for $|t|<<T$, as
$(-t)^{-\a_i-1} \sim (-\eta)^{-1}$. On the other hand, the horizon problem
of the standard cosmological model [50] is solved if, for every spatial
direction, the growth of the ratio $r^{i}(t)$ when $|t|$ is ranging from $T$ to
$t_1$,
is large enough to compensate
the  decreasing of the ratio in the subsequent decelerated phase down to
the present time $t_0$. This implies, in the hypothesis that the
pre-big-bang era is followed by the standard radiation-dominated (until
$t=t_2$) and matter-dominated evolution,
$$
\left (t_1\over T\right)^{-\a_i-1}\simeq \left (H_T\over H_1\right)^{-\a_i-1}
\Me \left(a_1\over a_0\right)\left(t_0\over t_1\right) =
$$
$$
=\left(t_1\over t_2\right)^{-1/2}\left(t_2\over t_0\right)^{-1/3}
=\left(H_1\over H_2\right)^{1/2}\left(H_2\over H_0\right)^{1/3} \simeq
10^{30}\sqrt{H_1\over M_p}\eqno(4.9)
$$
(the same condition is required to solve the flatness problem, see below).
For an expanding $d$-dimensional isotropic background $\a_i=1/\sqrt d$
(see eq.(2.48)), and the previous condition gives in particular, for
$t_1\simeq \la_s\simeq M_p^{-1}$,
$$
T\Me 10^{30\sqrt d/(\sqrt d+1)}\la_s,\,\,\,\,\,\,\,
H_T \me 10^{-30\sqrt d/(\sqrt d+1)}M_p \eqno(4.10)
$$
We shall thus assume that the scale $T$ appearing in the solution (4.3) is
much larger than the string scale $\la_s\simeq M_p^{-1}$.

This fact has an important consequence. In this case the background (4.3)
becomes in fact a good zeroth-order approximation to the general solution
of the full system of equations, consistent with the string equations of
motion not only in the asymptotic limit $t\ra -\infty$.

By adopting an iterative
approach, let us assume indeed the solution (4.3) to be a zeroth-order
approximation, and let us compute the first order corrections by inserting
that solution into the string equations of motion, in order to obtain the
corresponding value of $\ga_i(t)$. To this aim we observe that the given
background is characterized, asymptotically, by an accelerated metric with
shrinking event horizons (see eq.(4.6)). We recall that, in such a
background, the string equations of motion admit oscillating solutions,
corresponding to strings with constant proper size $L_s$, provided $L_s$
is smaller than the size of the event horizon $\sim H^{-1}(t)$ (stable
strings), while the solutions describe non oscillating strings with
$L_s(t)\sim a(t)$ if $L_s>H^{-1}$ (unstable strings) [27].

The evolution of a network of strings with some initial  distribution
in backgrounds of the type discussed above can be investigated [51].
 One can show that
the number $n(L_s,t)$ of strings (per unit length)
 of given size $L_s$, at
time $t$, must satisfy in the given background the approximate evolution
equation [51]
$$
{\pa n\over \pa t}=-H{\pa\over \pa L_s }[n
L_s\theta(L_s-H^{-1})]\eqno(4.11)
$$
where $\theta$ is the Heaviside step function.
 Its general solution can be
written in implicit form as [51]
$$
n(L_s,a(H))= n_0(L_s)\theta(H^{-1}-L_s)+ f\left(L_s\over a\right)\theta
(L_s-H^{-1}),
$$
$$
f\left(H^{-1}\over a\right)=n_0(H^{-1})\left(1+{\pa \ln a\over \pa \ln
H}\right )^{-1} \eqno(4.12)
$$
where $n_0$ is the initial string distribution.

The energy associated at a time $t$ with stable ($\rb_S$) and unstable
($\rb_U$) strings can be estimated as
$$
\rb_S \sim \int_{\la_s}^{H^{-1}}L_s n(L_s,t)dL_s,\,\,\,\,\,\,
\rb_U \sim \int^\infty _{H^{-1}}L_s n(L_s,t)dL_s
\eqno(4.13)
$$
However, for a perfect gas of stable strings $p_S=0$, while, for unstable
strings, $p_U=\pm\r_U/d$, with the sign fixed by the exponent $\a_i$ of
eq.(4.6), $sign\{ p_U\}= -sign \{\a_i\}$, as discussed in [27]. Therefore, the
ratio $\ga=p/\r$ as a function of time, for a perfect gas of strings in an
accelerated metric background, can be approximated as
$$
\ga (t)=\pm {1\over d} {\rb_U\over \rb_U+\rb_S} \eqno(4.14)
$$
By inserting into eqs.(4.13), (4.14) the solution (4.12) expressed for our
particular metric (4.6), with an initial
string distribution $n_0(L_s)\sim
L_s^{-3}$, one then finds for each spatial direction [51]
$$
\ga_i(t)=-{1\over d}\la_sH_i(t) ,\eqno(4.15)
$$
where $H_i$ is given by eq.(4.5). The above
  result is valid for $|H_i|<\la_s^{-1}
\simeq M_p$ and is not very sensitive to the initial string distribution.

We insert now this expression into the right-hand side of the field
equations (2.18), (2.19), by recalling that, for the pressureless background
(4.3) one has, to zeroth-order,
$$
\Ga_i^{(0)}= x_i={e^{\phi_0}\over 4L} t_i \eqno(4.16)
$$
Then, to next order,
$$
\Ga_i=x_i+\int _{-\infty}^x \ga_i(x^\pr)dx^\pr ={e^{\phi_0}\over 4L}
\left[t_i-{\la_s\over d}\int_{-\infty}^t {da_i\over a_i}\right]=
$$
$$
=\Ga_i^{(0)}\left[ 1-{\la_s\over dT}\ln \left (t-2T\over t\right)
\right]\eqno(4.17)
$$
According to our iterative approach, the integration of eqs.(2.18), (2.19) with
$D(x)$ determined by this new expression for $\Ga_i$ provides a first order
approximation to the background fields $a(t)$, $\phi(t)$.  The corrections to
the solution (4.3) due to a non-vanishing effective pressure of the string
gas are certainly negligible for $|t|>>T$, in the regime in which the
background (4.3) satisfies $H^2<<\r e^\phi \sim \dot{\fb}^2$. However, as
clearly shown by eq.(4.17), if $T>>\la_s$ then the first order corrections
keep small also in the $t\sim -T$ regime, in which $H^2\sim
\r e^\phi \sim \dot{\fb}^2$, and even in the limit $t\ra -t_1\simeq \la_s$,
in which $\r e^\phi <<\dot{\fb}^2 \sim H^2$. Within the assumption that $T$
is very large in string units, the solution (4.3) becomes then a good
approximation to the exact solution of the system of background equations
and string equations of motion, for the whole range $-\infty \leq t \leq
-t_1\simeq \la_s$.

We stress that, in this scenario, when $|t|<<T$ the source term
$\r e^\phi $ becomes negligible with respect to $H^2$ and $\dot{\fb}^2$
(see eq.(4.5)); quite independently of the exact value of the pressure and of
the particular type of equation of state at the scale $T$, the background
rapidly converges, for $|t|<<T$, to a phase of vacuum, dilaton-driven
accelerated evolution (as discussed in Sec.2), described by the metric (4.6).
We are left, therefore, with two phenomenological possibilities.

The first is the case in which $T$, and then the temporal extension of the
regime (4.6), is much larger than the minimal value fixed by eq.(4.9) to
secure a phenomenologically sufficient amount of inflation. This means, in
conformal time,
$$
|\eta_T|>>10^{30}\left(H_1\over M_p\right)^{1/2}
|\eta_1|\simeq |\eta _0|\eqno(4.18)
$$
where $\eta _0$ is the time when the largest scale, corresponding to the
minimum frequency mode $\om_0=H_0$, was pushed out of the event
horizon during the pre-big-bang phase. In this case, all  today's
observable scales crossed the horizon in the dilaton-driven regime (4.6), so
 that
the presently observed perturbation spectrum is wholly determined by the
metric behavior of that regime, quite independently of possible earlier
matter corrections to the background.

The second possibility is the case of nearly ``minimal" inflation,
corresponding to the equality in the condition (4.9), which implies then
$|\eta_T|\sim |\eta_0|$. In this case the largest scales crossed the horizon
when the contribution of the string sources to the metric background was
of the same order as the dilaton contribution. As a consequence, the low
frequency part of the scalar perturbation spectrum may be affected by the
matter corrections, and may be sensitive to  the particular type of equation
of state. The spectrum is thus to be computed by including the non-trivial
mixing induced by the source terms $T_{\mu\nu}$ and their perturbations,
$\da T_{\mu\nu}$, as discussed in Sec.3.

As anticipated in Sec.3, in this paper we will discuss the first possibility
only. We shall assume, in particular, that the phase of accelerated
evolution responsible for the solution of the standard kinematic problems,
and for the amplification of the perturbations (at all presently accessible
scales), is described by a three-dimensional, isotropic, dilaton-dominated
background with
$$
a(t)\sim (-t)^{-1/\sqrt 3},\,\,\,\,\,\,\, a(\eta)\sim (-\eta)^{-1/(\sqrt 3
+1)},
\,\,\,\,\,\,\, \phi \sim (3+\sqrt 3)\ln a,
$$
$$
t\leq -t_1 < 0 \,\,\,\,\,\,\,\,\,\,\,\,\,\, \eta \leq -\eta_1 <0 \eqno(4.19)
$$
(according to eqs.(2.48), (2.49)). More complicated scenarios, in particular
with higher-dimensional, anisotropic, sourceless backgrounds will be
analyzed elsewhere [41]).

The metric (4.19) describes superinflationary expansion [52]. In order to
obtain the dilaton spectrum, by applying eq.(3.32), we must transform
however the solution (4.19) into the E frame, where it takes the form (see
eqs.(2.57), (2.58))
$$
a_E(t_E)\sim (-t_E)^{1/3},\,\,\,\,\,\,\, a_E(\eta)\sim (-\eta)^{1/2},
\,\,\,\,\,\,\, \phi_E \sim -\sqrt {12}\ln a_E \eqno(4.20)
$$
This metric describes, for $t\ra 0_-$, a contracting background. Potentially,
this represents a difficulty of the whole scenario: indeed, the
approximation of a diluted string gas might be no longer valid in a
contracting background, as well as the approximated expression (3.36)
for the perturbation spectrum, obtained in the case of inflationary
expansion. Most important, it might seem impossible, in a contracting
background, to achieve a solution of the standard kinematic
problems [50],
thus rendering ``frame-dependent" the inflationary virtues of the
pre-big-bang scenario.

Surprisingly enough, however, this is not the case, as a consequence of the
fact that the contraction of the metric (4.20) is of the accelerated type,
with $\dot a <0$, $\ddot a <0$, $\dot H <0$ (one can show, in general, that all
the BD solutions describing superinflationary expansion, with or without
matter sources, are transformed through the Weyl rescaling (2.52) into E
backgrounds whose metric describes accelerated contraction [23]).  Let us
show, first of all, that a phase of accelerated contraction is equally good to
solve the kinematic problems of the standard model as a phase of
superinflationary expansion, characterized by
$\dot a >0$, $\ddot a >0$, $\dot H >0$. Consider indeed the so-called
flatness problem: the spatial curvature term becomes negligible with
respect to the other terms of the cosmological equations if the ratio
$$
r_1(t)={k\over a^2H^2} = {k\over \dot a^2} \eqno(4.21)
$$
goes to zero during inflation. Such a condition is certainly satisfied by a
metric which behaves, asymptotically, as
$$
a(t)\sim (-t)^\a, \,\,\,\,\,\,\,\, t<0,\,\,\,\,\,\,\, \a<1 \eqno (4.22)
$$
for $t\ra 0_-$. For $\a<0$ this metric parametrizes the known case of
pole-inflation (superinflationary expansion [52]). For $0<\a <1$ one has
instead accelerated contraction. In both cases the curvature scale is
growing, and $H,\dot H$ diverge as $t\ra 0_-$.

Accelerated contraction can also provide a solution to the horizon problem.
Indeed, by recalling the previous definition of particle (eq.(4.7)) and event
(eq.(4.8)) horizon, one finds that the ratio of their proper sizes in the
background (4.22),
$$
r_2(t) ={d_p(t)\over d_e(t)} \sim (-t)^{\a-1} , \eqno(4.23)
$$
diverges for $t\ra 0_-$. This means that causally connected regions will
always cross the horizon, asymptotically, not only in the case of
superinflationary expansion ($\a<0$), but also in the case of accelerated
contraction ($0<\a<1$).

It is important to stress that the condition for a successful resolution of
the horizon and flatness problems, when expressed in conformal time, is
exactly the same for superinflation and accelerated contraction. Quite
independently of $\a$, in fact, the ratio $r_2$ scales as $\eta^{-1}$, while
$r_1$ scales as $\eta^2$. The horizon problem is solved if $r_2$, evaluated at
the end of inflation ($\eta=\eta_f$), is larger than a present value of $r_2$
of order unity, rescaled up to $\eta_f$. Namely
$$
{|\eta_i|\over |\eta_f|} \Me {|\eta_0|\over |\eta_f|} \eqno(4.24)
$$
where $\eta_i$ denotes the beginning of the (contracting or expanding)
accelerated evolution (see also eq.(4.9)). The flatness problem is solved if
$r_1$, at $\eta=\eta_f$, is tuned to  a  small enough value, so that the
subsequent decelerated evolution leads to a present value of the ratio
$r_1(\eta_0)\me 1$. This implies
$$
\left (\eta_f\over \eta_i\right)^2 \me
\left (\eta_f\over \eta_0\right)^2 \eqno(4.25)
$$
which is clearly equivalent to the previous condition, as expected.

Therefore, if the accelerated phase of pre-big-bang is long enough to solve
the kinematic problems in the BD frame, where the metric describes
superinflationary expansion, then the solution holds also in the E frame
where the metric describes accelerated contraction, because the conditions
are the same in conformal time, and the conformal time is the same in the
two frames [23]. (We note, incidentally, that the kinematic problems can
thus be solved also if one chooses negative integration constants, $t_i<0$,
$\a_i<0$, for the background solution (4.6), corresponding to a metric
describing accelerated contraction already in the BD frame). For the
solution of the entropy problem, of course, a non-adiabatic phase
 associated to the inflation-radiation transition is necessarily required, in
addition to the accelerated kinematic, as recently stressed also in [53].

For what concerns the dilution of the string gas, we recall that in the BD
frame a model of source as a weakly interacting string network is a very
good approximation. In that frame, indeed, by starting at some initial time
$t_i$ with a packing factor = (average distance/average size) of order
unity, one ends up, at any subsequent time $t_f$, with a number of strings
per unit of string volume which is diluted as $n_f/n_i=(a_i/a_f)^d$ ($<1$
since the metric is expanding). In the E frame the metric is contracting, but
the string proper size $L_s^E(t)$ shrinks with time as
$L_s^E(t)=(a_E/a)\la_s$, where $a$ is the BD scale factor. As a
consequence, the number of strings per unit of string volume scales as
$n(t)=(L_s^E/\la_s)^d a_E^{-d}=a^{-d}$, and it is again diluted as time goes
up, exactly by the same amount as in the BD frame.
In other words, one finds that, at the end of inflation, a region of space
of initial linear dimensions $O(\la_s)$ has become exponentially large in
Planck units irrespectively of the frame one is using.
With similar arguments one can show [23] that the heating up of the string
gas with respect to the radiation, which is easy to understand in the BD
frame where the metric is expanding and the radiation is red-shifted, also
occurs in the E frame in spite of the fact that the radiation is blue-shifted
 because of the
contraction.

We want to show, finally, that the result (3.36) for the scalar perturbation
spectrum is also valid if the transformed metric, in the E frame, is a
contracting one. Consider in fact eqs.(3.33), (3.34), for the mode $v_k$ in
the non-oscillating regime. Since the variable $z=a\dot \phi/H$ goes like $a$
for $\phi \sim \b \ln a$, it might seem that for fast enough contraction the
second term of the expansion could dominate the first one, asymptotically,
thus changing the perturbation spectrum. We must recall, however, that
in the scenario that we are considering the universe evolves from an
initial phase of pre-big-bang to the standard, decelerated,
radiation-dominated expansion. In the BD frame the universe is always
expanding, $H_{BD}>0$, while in the E frame an initial contraction ($H_E<0$)
turns into a final expansion ($H_E>0$), with a necessary turning point of
$H_E$ at some time $\eta^\star$ near the transition time $-\eta_1$. On the
other hand, the conformal transformation (2.52) gives (in $d=3$)
$H_E=(H_{BD}-\dot \phi /2)e^{\phi/2}$: it follows that $\dot \phi \not= 0$
where $H_E=0$, and that $z \ra \infty$ for $\eta \ra \eta^\star$, so that the
first term of the expansion (3.33) is still the dominant one even in the E
frame.

By putting, in this frame, $a^\pr /a \simeq (a^\pr /a)_{\star}^\pr (\eta -
\eta^\star)$ for $\eta \ra \eta^\star$ ($z\ra \infty$), the amplification of
$\psi_k$ in this limit can thus be estimated as (using eq.(3.25))
$$
|\psi_k|\simeq |{\phi^\pr z\over k^2 a} \left(v_k\over z \right)^\pr |
\simeq |{(\phi_\star^\pr)^2 \over \sqrt{ k} (a^\pr /a)_\star^\pr}
\left(1\over
z\right)_{k\simeq aH} | \simeq {1\over \sqrt k} \left(1\over z \right)
_{k\simeq aH}
\eqno(4.26)
$$
where $\phi_\star^\pr \equiv \phi^\pr (\eta_\star)$, and we have used the
background equation $\phi^{\pr 2}$ = $-6 (a^\pr /a)^\pr$. We thus
recover for $|\da_\psi |^2$ = $k^3|\psi_k|^2$ in the E frame the standard
result (3.36), in spite of the contracting character of the transformed
pre-big-bang metric.

\vskip 2 cm

\centerline{\bf 5. Phenomenological constraints on the dilaton spectrum.}

In the simplified model of pre-big-bang  motivated and discussed
 in the previous Section, and
described by the background
(4.20), the dilaton
perturbations are amplified with a growing spectral distribution: one
has, from eq.(3.32),
$$
\Om_\chi (\om,t) \simeq GH_1^2\left (\om\over \om_1\right)^3
\left (H_1\over H\right)^2\left (a_1\over a\right )^4 \eqno(5.1)
$$
The total dilaton energy density $\r_\chi(t)$ is thus dominated by the
highest frequency mode $\om_1$,
$$
\Om_\chi (t) ={\r_\chi(t) \over \r_c(t)}=
\int^{\om_1} {d\om \over \om} \Om_\chi(\om ,t)
\simeq GH_1^2
\left (H_1\over H\right)^2\left (a_1\over a\right )^4 \eqno(5.2)
$$
and since $\r_\chi(t)$ decreases in time like the radiation density ($\sim
a^{-4}$), its value in critical units, $\Om_\chi (t)$, remains constant
during the radiation-dominated era ($H\sim a^{-2}$). The requirements
that the produced dilatons do not overclose the universe ($\Om_\chi <1$)
in the radiation era thus imposes the condition
$$
H_1 \me M_p\eqno(5.3)
$$
which is also needed from a similar constraint on gravitons (this
constraint is valid not only for
the particular case (5.1) but also, more generally, for all growing dilaton
spectra, whose integration leads to an $\Om_\chi (t)$ similar  to that of
eq.(5.2)).

If the dilaton would be massless, this would be the end of the story.
However, in spite of some recent attempt [54] trying to motivate the
possibility of a massless dilaton in a string theory context, present
conventional wisdom seems to favor a non-vanishing dilaton mass, with a
mass value closely related, in particular, to the phenomenology of
supersymmetry breaking (see [55] for a recent discussion). But, even
independently from possible supersymmetry motivations, a
non-vanishing mass seems to be a compulsory consequence of the fact that
dilatons couple non-universally to macroscopic matter, with coupling
strength larger than (or at least equal to) the gravitational one [4].
We shall reproduce below, for completeness, the argument given in
ref. [4] making it actually slightly more general.

The large distance behavior of dilaton couplings is determined indeed by
the string effective action $\Ga$, whose general form, including possible
loop corrections, can be written as
$$
\Ga= -\int d^Dx\sqrt{|G|} \left [ Z_R(\phi) R(G)+Z_\phi (\phi) G^{\mu\nu}
\pa_\mu \phi \pa_\nu \phi+ V(\phi) \right ] +\Ga_M,
$$
$$
\Ga_M= \sum_i\int d^Dx\sqrt{|G|}\left [{1\over 2}Z_k^{i}(\phi)G^{\mu\nu}
\pa_\mu \psi _i\pa_\nu \psi_i +Z_m^{i} (\phi)\psi_i^2 +
interaction\,\,\,\, terms\right] \eqno(5.4)
$$
Here $Z_R, Z_\phi, Z_k, Z_m$ are complicated (known in principle,
but unknown in practice) coupling functions,
$G_{\mu\nu}$ is the (dimensional) sigma-model metric, and we have
represented the matter part of the action as a set of (dimensionless)
scalar fields $\psi_i$ (fundamental fermions can be added without
difficulty [4]). In order to evaluate the effective dilaton couplings
we must restore, first of all, the canonical form of the kinetic energy
terms, by rescaling field and masses. The dilaton coupling to matter fields
is then obtained from the effective interaction Lagrangian, expressed in
terms of the rescaled variables.

To this aim we note that the graviton kinetic term of eq.(5.4) reduces to
the canonical Einstein Lagrangian by putting
$$
G_{\mu\nu}=g_{\mu\nu} M_p^2Z_R^{-2/(d-1)}\eqno(5.5)
$$
where $g_{\mu\nu}$ is the dimensionless Einstein metric. By defining a
new scalar field (with canonical dimensions) $\sg$, such that
$$
{d\sg\over d\phi} =M_p^{(d-1)/2} \left [{2d\over d-1}\left(d\ln Z_R \over
d\phi\right)^2 -2{Z_\phi\over Z_R}\right]^{1/2}\eqno(5.6)
$$
and new rescaled matter fields
$$
\hat \psi_i=M_p^{(d-1)/2}\sqrt{Z_k^{i}(\sg) \over Z_R(\sg)}\,\,
\psi_i\eqno(5.7)
$$
the action (5.4) can be written in canonical form as
$$
\eqalign{
\Ga= -\int d^Dx\sqrt{|g|} [ - &{R(g)\over 16 \pi G_D}+{1\over 2}
g^{\mu\nu} \pa_\mu \sg \pa_\nu \sg-W(\sg) + \cr
&+ \sum_iJ\left ({1\over 2}g^{\mu\nu}
\pa_\mu \hat\psi _i\pa_\nu \hat\psi_i +{1\over
2}\mu_i^2(\sg)\hat\psi_i^2 \right) ] \cr}\eqno(5.8)
$$
where $16\pi G_D=M_p^{1-d}$, $W=VM_p^{d+1}Z_R^{-2/(d-1)}$ and
$$
\mu_i^2(\sg) ={2M_p^2 Z^{i}_m(\sg) \over Z^{i}_k(\sg) Z^{2/(d-1)}_R(\sg)}
\eqno(5.9)
$$

We expand now the effective matter-dilaton interaction Lagrangian,
$\mu_i^2\hat \psi_i^2$, around the value of $\sg$ which extremizes the
dilaton potential (and which can always be assumed to coincide with
$\sg=0$, after a trivial shift). Defining
$$
{1\over 2} \mu_i^2(\sg)\hat\psi_i^2 = {1\over 2} m_i^2\hat\psi_i^2
+ g_i\sg \hat\psi_i^2 +
O(\sg^2) ~, ~\eqno(5.10)
$$
we can express the rescaled mass of the matter fields as
$$
m_i^2=\left [\mu_i^2(\sg)\right]_{\sg=0}=
2M_p^2 \left( Z^{i}_m\over Z^{i}_kZ^{2/(d-1)}_R\right)_{\sg=0}
\eqno(5.11)
$$
and the effective matter-dilaton couplings (including loop corrections) as
$$
g_i= {1\over 2}\left(d\mu_i^2\over d\sg \right)_{\sg=0}
={1\over 2}\left({d\phi\over d\sg}{d\mu_i^2\over d\phi}\right)_{\sg=0}=
$$
$$
={m_i^2\over 2 M_p^{(d-1)/2} }\left [{2d\over d-1}\left(d\ln Z_R \over
d\phi\right)^2 -2{Z_\phi\over Z_R}\right]^{-1/2}_{\phi=0}
\left [{d\over d\phi} \ln \left(\mu_i^2(\phi)\over
M_p^2\right)\right]_{\phi=0}\eqno(5.12)
$$

In the weak coupling regime $Z_R=Z_\phi=e^{-\phi}=g^{-2}$, where $g$
is the gauge coupling constant of the superstring unification group, and
one finds that the effective dilaton coupling strength $g_i/m_i$ deviates
from the standard ``gravitational charge"
 $\sqrt{4\pi G_D}m_i$ by the factor
$$
k_i= {g_i\over m_i^2 \sqrt{4\pi G_D}} \simeq
\left(d-1\over 2\right)^{1/2}
 \left [g^2 {\pa \over \pa
g^2} \ln \left(\mu_i\over M_p\right)^2\right]_{\phi=0} \simeq
$$
$$
\simeq
1+\left[{\pa\over \pa \phi} \ln \left(Z_m^{i}\over
Z_k^{i}\right)\right]_{\phi=0}\eqno(5.13)
$$
where the last equality refers to $d=3$.
As already stressed in Ref.[4], eqs.(5.12) and (5.13) clearly show two
important phenomenological effects. The first is that string loop
contributions violate the universality of the effective dilaton couplings, as
the factors $k_i$ being different for particles whose mass has different
 origins; the
second is that the dilaton coupling is even stronger than the graviton
coupling, as $k_i\geq 1$ for all known cases [4] and, in particular, for
the confinement-generated component of hadronic masses. Such a conclusion
has been recently challenged [54] on the basis of a possible new
mechanism forcing eq.(5.13) to give a vanishing (or very small)
result. It is difficult for us to understand how such a cancellation
can occur for any realistic present value of the dilaton.

The existence of
 a non-universal scalar
force of gravitational strength may be reconciled with the
Eotvos-Dicke-Braginski experiments only if its range is finite [56]. By
considering the present results obtained from tests of the equivalence
principle [5], it follows in particular
 that the dilaton corrections to low
energy Newtonian gravity are only allowed if their range is smaller than
about $1\,cm$, namely for a dilaton mass
$$
m\Me m_0 =10^{-4}\, eV \eqno(5.14)
$$

We now turn to the discussion of bounds following from the energy
stored in the dilaton perturbations (or, if we prefer, in actual
scalar particles associated with that field). Our discussion
follows closely that of ref. [8].
We recall now that the expression (5.2) for the dilaton energy density
was obtained neglecting the contribution of the rest energy to the proper
oscillation frequency $E(t)=\sqrt{(k/a)^2+m^2}$ (see eq.(3.27)). Therefore,
for a massive dilaton, eq.(5.2) is only valid at early enough times such
that $\om_1(t)=k_1/a >m$ (and thus it is certainly not valid today, in view
of the previous bound and of eq.(1.4)). At later times eq.(5.2) is to be
corrected to take into account the mass contribution. Let us suppose,
first of all, that $m<H_1=\om_1(t_1)$, so that eq.(5.2) holds initially, at
the beginning of the radiation era, and that $\Om_\c$ stays dominated by
the highest mode $\om_1$ also in the non-relativistic regime.
The proper frequency $\om_1$ is
red-shifted as the curvature scale decreases in time, and the dominant
mode becomes non-relativistic at a scale $H_{nr} \equiv H(t_{nr})$ such
that
$$
\om_1(t_{nr})= H_1 {a_1\over a_{nr}}=m \eqno(5.15)
$$
For $H\me H_{nr}$ the oscillation frequency $E_1(t)=\sqrt{\om_1^2(t)
+m^2}$ is dominated by the mass contribution, and the corrected dilaton
energy density can be obtained from eq.(5.2) through the rescaling
$$
\Om_\chi (t) \ra {m\over \om_1}
 \Om_\chi(t)
\simeq GmH_1
\left (H_1\over H\right)^2\left (a_1\over a\right )^3\equiv G{m^4\over
H^2} \left (a_{nr}\over a \right )^3 \eqno (5.16)
$$
(see also [8]). The cosmological bounds on $\Om_\chi$ become then
bounds on the dilaton mass, and provide us with a phenomenologically
allowed region in the $(m,H_1)$ parameter space.

Consider, first of all, the critical density bound $\Om_\chi <1$. Once that
the condition (5.3) is satisfied, the dilaton density $\r_\chi$ is certainly
subcritical (and constant with respect to the radiation density $\r_\ga$)
for all scales $H>H_{nr}$. When the scale drops below $H_{nr}$, however,
the dilaton density becomes non-relativistic according to eq.(5.16), the
ratio $\r_\chi/\r_\ga$ begins to grow in time like the scale factor, and
the equality $\r_\chi=\r_\ga$ may be reached at some initial scale
$H_i\equiv H(t_i)$. We have now various phenomenological possibilities,
depending on the value of $H_i$ and $H_{nr}$.

Suppose first that the transition to the non-relativistic regime occurs
when the universe is already matter-dominated, namely for $H_{nr}\me
H_2$ (where $H_2 \sim 10^{-27} eV$ is the scale of radiation-matter
transition) which means, according to the definition (5.15)
$$
m \me \sqrt{H_1H_2} \eqno(5.17)
$$
In the matter-dominated era, the dilaton energy density (in critical units)
(5.16) stays fixed at the value
$$
\Om_\chi \simeq GH_1^2{m\over \sqrt{H_1H_2}}\eqno(5.18)
$$
The requirement $\Om_\chi<1$ is thus automatically satisfied, in this
case, because of eqs.(5.9), (5.17). The same happens if $H_{nr}>H_2$, but
the scale $H_i$ of dilaton-radiation equilibrium belongs to the epoch of
matter domination, $H_i<H_2$. In this case, indeed, $H_{nr}=m^2/H_1$,
and $H_i$ is fixed (according to eq.(5.16)) by
$$
{\r_\chi(t_i)\over \r_\ga(t_i)} ={\r_\chi(t_{nr})\over \r_\ga(t_{nr})}
{a_i\over a_{nr}} \simeq GH_1^2{m\over \sqrt{H_1H_2}}\left (H_2\over
H_i \right )^{2/3} =1\eqno(5.19)
$$
The condition $H_i<H_2$, reading now
$m \me \sqrt{H_1H_2} (GH_1^2)^{-1}$,
again implies $\Om _\chi<1$, where $\Om_\chi$ is
the constant dilaton density (5.18).

Dilatons are always subdominant even if $H_{nr}>H_i>H_2$, but the
decay scale $H_d$ (1.7) is larger than $H_i$, so that dilatons are forced to
dissipate their coherent energy density, converting it into radiation
before becoming dominant. In this case we have, from eq.(5.16),
$$
H_i= H_{nr}\left [\r_\chi(t_{nr})\over \r_\ga(t_{nr}) \right]^2
={m^2H_1^3\over M_p^4} <H_{nr}\eqno(5.20)
$$
and the condition $H_d>H_i$ reads
$$
m > H_1 \left (H_1\over M_p\right)^2\eqno(5.21)
$$
If, on the contrary, $H_{nr}>H_i>H_2$ and $H_i>H_d$, then in order
 to avoid
contradiction with the presently observed density of non-relativistic
matter it is necessary, first of all,
to impose that dilatons already decayed, $H_d>H_0$. This gives
$m^3 > M_p^2 H_0$ (numerically, $m> 100~ MeV$). This is not sufficient
though and one is
left, in this case, with two possible alternatives.

The first is the one in which the reheating temperature $T_r$
associated with dilaton decay
$$
T_r\simeq \sqrt{M_pH_d} \simeq \left (m^3\over
M_p\right)^{1/2}\eqno(5.22)
$$
is smaller than the temperature scale $T_N\sim 1MeV$ required by
nucleosynthesis. This provides
$$
m\me 10^4\, GeV \eqno(5.23)
$$
In such case we must impose that nucleosynthesis occurred before
dilaton dominance, $H_N\simeq (1\, MeV)^2/M_p>H_i$, and that the entropy
$\Da S$ associated with dilaton decay is small enough, in order not to destroy
all light nuclei already formed. The temperature $T_d$ of the radiation gas
already present at the scale $H_d$ is in fact, from eq.(5.20),
$$
T_d=T_i\left(a_i\over a_d\right)\simeq (M_pH_i)^{1/2}\left(H_d\over
H_i\right)^{2/3}\simeq \left(m^{10}\over
M_pH_1^3\right)^{1/6}\eqno(5.24)
$$
The reheating of the radiation gas from $T_d$ to $T_r$ thus produces an
entropy increase
$$
\Da S \simeq \left (T_r\over T_d\right)^3\simeq \left(H_1^{3}\over
mM_p^2\right)^{1/2}\eqno(5.25)
$$
By imposing $\Da S \me 10$ in order to preserve nucleosynthesis [57] one
obtains the bound
$$
m\Me 10^{-2} {H_1^3\over M_p^2}\eqno(5.26)
$$

The second alternative corresponds to $T_r\Me T_N$, i.e. $m \Me
10^4\,GeV$, and allows then a nucleosynthesis phase subsequent to
dilaton decay. In such case, the only phenomenological constraint is
possibly imposed by primordial baryogenesis. The maximum tolerable
amount of entropy, in order not to wash-out any pre-existing
baryon-antibaryon asymmetry, is somewhat model-dependent, but in
general $\Da S \me 10^{5}$ seems to be acceptable [57,58]. This implies
the bound
$$
m\Me 10^{-10} {H_1^3\over M_p^2}\eqno(5.27)
$$
Note, however, that this last condition may be evaded in the case of
low energy baryogenesis and, in particular, in the case of baryogenesis
associated with the dilaton decay itself [55,58], occurring at scales not
much distant from nucleosynthesis.

The bounds so far considered refer to the case in which the dilaton
energy density stays always dominated by the contribution of the highest
frequency mode $\om_1$, even in the non-relativistic regime. This is
certainly true for the distribution (5.1), with spectral index $\da=3$,
but for a more complete phenomenological discussion let us consider also
\footnote{*}{The necessity of considering lower modes and of
distinguishing
two intervals in $\da$
was pointed out to us by A. Starobinsky.}
the coherent oscillations, with frequency $m$, of the lower modes with
$\om(t_m)\me m$, which begin at the scale $H_m\equiv H(t_m)=m$, when the
mode $\om_1$ is still relativistic ($m>H_{nr}$). For $H\me m$ such modes
provide a non-relativistic contribution to the dilaton energy density
which can be written (for a general spectrum with $\da>0$) as
$$
\Om_\c(t)\simeq
Gm^2\left(H_1\over H\right)^2\left(m\over H_1\right)^{\da \over 2}
\left(a_m\over a\right)^3
=Gm^2\left(H_1\over H\right)^2\left(m\over H_1\right)^{\da-3 \over 2}
\left(a_1\over a\right)^3
\eqno(5.28)
$$
This contribution is initially negligible with respect to the
relativistic part of the dilaton energy density, $\sim GH_1^2$,
dominated by $\om_1$. However, during the radiation era it grows in time
with respect to $GH_1^2$, and it may dominate the total dilaton energy
if the equality
$$
Gm^2\left(H_1\over H\right)^2\left(m\over H_1\right)^{\da \over 2}
\left(H\over m\right)^{3/2} = GH_1^2
\eqno(5.29)
$$
occurs at a scale $H>H_{nr}=m^2/H_1$, namely for $(m/H_1)^{\da-1}>1$,
which implies $\da <1$.

The previous bounds, obtained in the hypothesis of $\om_1$-dominance,
are thus valid for all growing spectra with $\da \geq 1$. For lower
spectral slopes, $0<\da<1$, the dominant contribution to $\Om_\c$ (the
one to be bounded) becomes that of eq.(5.28) and, as a consequence, the
bounds acquire a dependence on $\da$. Indeed, let $H_1$ be the scale
marking the equality $\r_\c =\r_\ga$. Then the condition $\Om_\c <1$,
with $\Om_\c$ given by eq.(5.28) and $0<\da<1$, is always satisfied for
$H_i<H_2$, which means
$$
m<\left(H_2M_p^4H_1^{\da-4}\right)^{1/(\da+1)} \eqno(5.30)
$$
It is also satisfied if $H_i>H_2$, but $H_d>H_i$, which means
$$
m>\left(H_1^{4-\da}M_p^{-2}\right)^{1/(2-\da)}
\eqno(5.31)
$$
If, however, $H_i>H_2$ and $H_i>H_d$, then the dilatons must have
already decayed, $H_d>H_0$. Their decay generates an entropy
$$
\Da S =\left(T_r\over T_d\right)^3=
\left(H_1^{4-\da}m^{\da-2} \over M_p^2\right)^{1/2}\eqno(5.32)
$$
If $m<10^4~GeV$ the reheating temperature is too low to allow
nucleosynthesis: we must impose that nucleosynthesis already occurred,
$H_i<H_N$, and that [57] $\Da S \me 10$, which means
$$
m\me \left(10^{-2}M_p^{-2}H_1^{4-\da}\right)^{1/(2-\da)}\eqno(5.33)
$$
If, on the contrary, $m>10^4~GeV$, the nucleosynthesis scale is
subsequent to dilaton decay, and the only possible constraint [57,58] is
$\Da S\me 10^{-5}$, namely
$$
m\me \left(10^{-10}M_p^{-2}H_1^{4-\da}\right)^{1/(2-\da)}\eqno(5.34)
$$

This complete the compilation of the phenomenological bounds for a
growing dilaton spectra, $\da>0$, with $m<H_1$.
Let us consider also, for the sake of
completeness, the ''heavy
dilaton" case, $m>H_1$, although this case is very unnatural in a string
theory context where $H_1$ is expected to be close to the Planck scale.

As discussed in Sec.3, in the radiation phase ($H<H_1, a\sim \eta$) the
dilaton modes $\chi_k$ decouple from the other scalar fluctuations, and
satisfy the free equation (which includes in general the mass contribution)
$$
\overline \chi_k^{\se}+ (k^2+m^2a^2)\overline \chi_k =0
,\,\,\,\,\,\,\,\,\,\,\,\, \overline \chi_k = a \chi_k \eqno(5.35)
$$
For all modes that are relativistic at the beginning of the radiation era
($\eta=-\eta_1$), namely for $k>ma_1$, one then recover the dilaton
spectrum (3.32) by matching the pre-big-bang solution (3.30) with the
plane wave (3.27), which is solution of eq.(5.35) in the case of negligible
mass. If, however,
$$
m>H_1=\om_1(t_1)={k_1\over a_1} \eqno(5.36)
$$
then the mass term cannot be neglected even in the case of the highest
mode $k_1$. All modes are non-relativistic already at $\eta=-\eta_1$,
and satisfy the approximate equation
$$
\overline \chi_k^{\se}+ m^2a^2\overline \chi_k =0
\eqno(5.37)
$$
By using the identity $m^2 a^2= m^2H_1^2a_1^4\eta^2$, valid in the
radiation era, the general solution of (5.37), for $mH_1a_1^2\eta^2\simeq
(m/H_1)(\eta/\eta_1)^2>1$ can be expressed as
$$
\overline \chi ={1\over \sqrt{mH_1a_1^2\eta}}
(c_+e^{-i mH_1a_1^2\eta^2} + c_-e^{imH_1a_1^2\eta^2})\eqno(5.38)
$$
The matching of this solution with the Bessel solution (3.30) gives then
the Bogoliubov coefficient for the case $m>H_1$
$$
|c_-(k)|^2 \simeq {m\over H_1} (k\eta_1)^{-2|\nu|}\eqno(5.39)
$$
and the corresponding non-relativistic dilaton spectrum (in critical units)
$$
\Om_\chi(\om,t)\simeq {G\om\over H^2}{d\r_\chi\over d\om} \simeq
G{m\om^3\over H^2}|c_-|^2\simeq
$$
$$\simeq Gm^2\left (\om\over \om_1\right)^{2-2\a}
\left (H_1\over H\right)^2\left (a_1\over a\right )^3\eqno(5.40)
$$
(we
recall that  $2-2\a=3$ for the particular model of pre-big-bang that we
have considered).

For a relic background of massive dilatons with $m>H_1$, eq.(5.2) is thus
to be replaced by the non-relativistic energy density
$$
\Om_\chi (t) =
\int^{\om_1} {d\om \over \om} \Om_\chi(\om ,t)
\simeq Gm^2
\left (H_1\over H\right)^2\left (a_1\over a\right )^3 \eqno(5.41)
$$
One must thus impose the bound
$$
m\me M_p\eqno(5.42)
$$
to avoid an over-critical density of massive dilatons at the beginning of
the radiation era. In this case there are no further bounds on $m$ since,
as a consequence of eq.(5.42), the dilaton energy is dissipated before a
possible dominance. Indeed, the initial scale $H_i$ corresponding to
$\Om_\chi =1$ is, from eq.(5.41),
$$
H_i=H_1\left (m\over M_p\right)^4 .\eqno(5.43)
$$
Therefore, $H_i/H_d=H_1m/M_p^2<1$, just because of eq.(5.42) (unless
$H_1>M_p$, but this is to be excluded to avoid over-critical dominance of
other massless particles, such as gravitons, produced by the same
background transition).

By collecting all previous phenomenological constraints we obtain a final
allowed region in the ($m,H_1$) plane, which depends on $\da$ and which
is illustrated in {\bf Fig.2} for the three cases $\da=0$, $\da=0.5$
and $\da=1$. We recall that for large enough spectral slope, $\da>1$,
the bounds become slope-independent and coincide with those of the
$\da=1$ case, which defines then the maximum allowed region for a
growing dilaton spectrum (as stressed also in [8]). As clearly shown by
the pictures, one of the main effects of a positive spectral index is
that light, not yet decayed dilatons become compatible with higher and
higher inflation scales as $\da$ ranges from $0$ to $1$. If we take, for
instance, $H_1\simeq 10^{-5}~M_p$
 as a typical reference value, we find
allowed mass windows for a presently dominating background of relic
dilatons for all spectra with $\da \Me 0.1$. For $\da \Me 1$, moreover,
even the $TeV$ mass range, which is supported by some supersymmetry
breaking arguments [55], but which lies in the most
unfavorable region for the various cosmological constraints,
  may become compatible with $H_1\simeq 10^{-5}~M_p$.

We note, finally, that our allowed regions refer to the cosmological
amplification of the quantum fluctuations of the dilaton background. The
classical dilaton background is assumed here to  sit at the minimum of
the dilaton potential, with negligible (with respect to quantum
fluctuations) oscillations around it. The classical oscillation amplitude,
however, could be too large to be negligible. In that case one should add,
to the bounds discussed here, the bounds on $m$ obtained by taking into
account the contribution of the classical oscillations to the total
cosmological energy density [7]. The initial amplitude of such possible
classical oscillations depends, however, on the details of the transition
from the accelerated to the decelerated regime.
Having neglected such possible additional
bounds, the allowed region determined here is to be regarded, for
 each value of $\da$, as the
{\it maximal} allowed region in parameter
space.

\vskip 2 cm

\centerline{\bf 6. Conclusions.}

In this paper we have presented the general solution of the equations
obtained from the low energy string effective action, for the case of
space-independent background fields, vanishing dilaton potential and
classical strings as possible matter sources. In the perfect fluid
approximation, the solution is uniquely fixed by the choice of the
equation of state. We have shown that a model of initial sources such as a
pressureless gas of weakly interacting strings provides an
approximate, but consistent solution to the full system of background
equations and string equations of motion.

This model supports a scenario in which an initial flat perturbative
string vacuum evolves towards a high curvature, strong coupling
regime through a phase of accelerated expansion or contraction. This is
the so-called pre-big-bang epoch, originally motivated in [2,25,27] by
the duality symmetries of the string effective action. Accelerated
contraction, typical of an Einstein frame representation of the
pre-big-bang scenario, works as well as the more conventional
inflationary kinematic (accelerated expansion) in order to solve the
standard cosmological problems.

We have derived, for an isotropic background, the general coupled
system of scalar (metric plus dilaton) perturbation equations, and shown
that the transition from the pre-big-bang phase to the standard,
radiation-dominated cosmology, is associated with a copious production of
cosmic dilatons, whose spectral distribution grows with frequency. We
have discussed the consequent phenomenological bounds on the dilaton
mass and on the inflation scale, by combining it with other bounds
obtained from tests of the equivalence principle. As a result, we have
found  allowed windows for the dilaton mass compatible with a
possible ``dilatonic" solution of the dark matter problem.

The particular model of pre-big-bang considered in this paper leads, however,
to scalar metric perturbations which cannot be taken as the origin of
the observed CMBR anisotropy, since their spectrum grows too fast
with frequency. This can be a welcomed result to people believing
 in a different source of anisotropy (e.g. in
cosmic strings, which could naturally arise in this context as remnants of
the violent, non-adiabatic transition from the growing to the decreasing
curvature regime). On the other hand, such a result
 cannot be really taken as typical of
our scenario, since we have neglected various effects which
 may decouple metric and dilaton perturbations leading to a
different (possibly flatter) spectrum of metric perturbations.
 We have in mind, for
instance, the contribution of matter sources to the background solutions
during the phase of parametric amplification of the vacuum fluctuations
 and/or
 a non-trivial evolution of the internal dimensions, whose
dynamical compactification
introduces at least one additional scalar variables in the perturbations
equations.

The most serious omission of this paper is, however, the lack of a
detailed description of the transition between the pre-big-bang regime
(growing dilaton and curvature scale) and the post-big-bang regime
(constant dilaton and decreasing curvature scale). We avoided to face
this problem in this paper, as we believe that the low energy effective
action adopted here cannot represent any longer an adequate
approximation for that purpose. A recent investigation [59]
strongly suggests that
higher curvature corrections should play a fundamental
role in solving the ``gracious exit" problem in string cosmology.
Furthermore a non-perturbative dilaton potential must be added in the
post-big-bang era in order to pin down the dilaton to its present value
and to give it a mass.

We do not conceal that the high-curvature transition from
inflation to standard cosmology is, at present, the least understood
aspect of the whole string cosmology scenario and that it certainly deserves
future detailed investigations. We stress  however  that, once the
mechanism which stops the growth of the dilaton and of the curvature and
 converts
non-adiabatically their kinetic energy into radiation is
clarified,  the phenomenology developed in this paper should remain
 valid without
further modifications and quite independently of the details of the transition
process.

\vskip 2 cm
\noi
{\bf Acknowledgments.}
We are grateful to R. Brandenberger, R. Brustein, M. Giovannini, K.
Meissner, V. Mukhanov, R. Ricci and A. Starobinski for many helpful
discussions.
G. V. wishes to thank A. Linde for a useful correspondence. M. G. wishes
to thank the CERN Theory Division for hospitality and financial support
during part of this work.

\vfill\eject
\centerline{\bf References}
\vskip 0.5 cm
\item{1.}L. P. Grishchuk, Sov. Phys. JEPT 40, 409 (1975);

A. A. Starobinski, JEPT Lett. 30, 682 (1979);

V. A. Rubakov, M. Sazhin and A. Veryaskin, Phys. Lett. B115, 189 (1982);

R. Fabbri and M. Pollock, Phys. Lett. B125, 445 (1983);

L. Abbott and M. Wise, Nucl. Phys. B244, 541 (1984)

\item{2.}M. Gasperini and G. Veneziano, Astropart. Phys. 1, 317 (1993)

\item{3.}M. Gasperini and M. Giovannini, Phys. Rev. D47, 1529 (1993)

\item{4.}T. R. Taylor and G. Veneziano, Phys. Lett. B213, 459 (1988);

 \item{5.}See for instance E. Fischbach and C. Talmadge, Nature 356, 207
(1992)

\item{6.}J. Ellis et al., Phys. Lett. B228, 264 (1989).

\item{7.}G. D. Coughlan et al., Phys. Lett. B131, 59 (1983);

J. Ellis, D. V. Nanopoulos and M. Quiros, Phys. Lett. B174,176 (1986);

J. Ellis, N. C. Tsamish and M. Voloshin, Phys. Lett. B184, 291 (1986);

B. De Carlos et al., Phys. Lett. B318, 447 (1993)

\item{8.}M. Gasperini, Relaxed bound on the dilaton mass in a string
cosmology scenario, DFTT-03/94 (January 1994), gr-qc/9401026

\item{9.}A. S. Goncharov, A. D. Linde and M I. Vysotsky, Phys. Lett. B147,
279 (1984)

\item{10.}G. Smoot et al., Astrophys. J. 396, L1 (1992)

\item{11.}C. L. Bennet et al., Cosmic temperature fluctuations from two

years of COBE DMR observations,
 astro-ph/9401012 (January 1994)

\item{12.}T. Piran et al., Limits on the primordial fluctuation
spectrum,
void sizes and CMBR anisotropy, astro-ph/9305019

\item{13.}A. D. Linde, Phys. Lett. B259, 38 (1991);

Phys. Rev. D49 (1994) (in press)

\item{14.}E. J. Copeland et al., False vacuum inflation with Einstein
gravity, SUSSEX-AST 94/1-1 (January 1994), astro-ph/9401011

\item{15.}S. Mollerach, S. Matarrese and F. Lucchin, Blue perturbation
spectra from inflation, astro-ph/9309054

\item{16.}J. Mc Donald, Phys. Rev. D48, 2468 (1993)

\item{17.}E. W. Kolb, D. Salopek and M. S. Turner, Phys. Rev. D42, 3925 (1990)

\item{18.}M. S. Turner and L. M. Widrow, Phys. Rev. D37, 3428 (1988)

\item{19.}C. Lovelace, Phys. Lett. B135, 75 (1984);

E. S. Fradkin and A. A. Tseytlin, Nucl. Phys. B261, 1 (1985);

C. G. Callan, D. Friedan, E. J. Martinec and M. J. Perry, Nucl. Phys.
B262,

593 (1985)

\item{20.}M. Gasperini and G. Veneziano, Phys. Lett. B277, 256 (1992)

\item{21.}K. A. Meissner and G. Veneziano, Phys. Lett. B267, 33 (1991)

\item{22.}K. A. Meissner and G. Veneziano, Mod. Phys. Lett. A6, 3397
(1991)

\item{23.}M. Gasperini and G. Veneziano, Mod. Phys. Lett. A8, 3701
(1993)

\item{24.}M.Mueller, Nucl. Phys. B337, 37 (1990)

\item{25.}G. Veneziano, Phys. Lett. B265, 287 (1991)

\item{26.}A. A. Tseytlin, Mod. Phys. Lett. A6, 1721 (1991)

\item{27.}M. Gasperini, N. Sanchez and G. Veneziano, Nucl. Phys. B364,
365 (1991);

Int. J. Mod. Phys. A6, 3853 (1991)

\item{28.}J. Bardeen, Phys. Rev. D22, 1822 (1980)

\item{29.}M. Sasaki, Prog. Theor. Phys. 70, 394 (1983)

\item{30.}V. Mukhanov, H. A. Feldman and R. Brandenberger,
Phys. Rep. 215, 203 (1992)

\item{31.}S. Mollerach and S. Matarrese, Phys. Rev. D45, 1961 (1992)

\item{32.}N. Deruelle, C. Gundlach and D. Langlois,
Phys. Rev. D46, 5337 (1992)

\item{33.}N. Deruelle, C. Gundlach and D. Polarski, Class. Quantum Grav.
9, 137 (1992)

\item{34.}L. P. Grishchuk, in Proc. of the 6th Marcel Grossmann Meeting
(Kyoto, 1991), ed. by H. Sato (World Scientific, Singapore, 1992)

\item{35.}M. Sasaki, Prog. Theor. Phys. 76, 1036 (1986)

\item{36.}V. F. Mukhanov, Sov. Phys. JEPT 67, 1297 (1988)

\item{37.}E. D. Stewart and D. H. Lyth, Phys. Lett. B302, 171 (1993)

\item{38.}V. N. Lukash, Sov. Phys. JEPT 52, 807 (1980)

\item{39.}G. V. Chibisov and V. N. Mukhanov, Mon. Not. R. Astron. Soc.
200, 535 (1982)

\item{40.}N. Deruelle, C. Gundlach and D. Langlois, Phys. Rev. D45,
R3301 (1992)

\item{41.}M. Gasperini and G. Veneziano, in preparation

\item{42.}N. D. Birrel and P. C. W. Davies, Quantum fields in curved
space (Cambridge University Press, Cambridge, England, 1982)

\item{43.}F. Lucchin and S. Matarrese, Phys. Lett. B164, 282 (1985);

B. Allen, Phys. Rev. D37, 2078 (1988);

V. Sahni, Phys. Rev. D42, 435 (1990);

L. P. Grishchuk and M. Solokhin, Phys. Rev. D43, 2566 (1991);

M. Gasperini and M. Giovannini, Phys. Lett. B282, 36 (1992)

\item{44.}L. P. Grishchuk and Y. V. Sidorov, Phys. Rev. D42,
3413 (1990);

L. P. Grishchuk, Phys. Rev. Lett. 70, 2371 (1993)

\item{45.}M. Gasperini and M. Giovannini, Phys. Lett. B301, 334 (1993)

\item{46.}M. Gasperini, M. Giovannini and G. Veneziano,
Phys. Rev. D48, R439 (1993)

\item{47.}R. Brandenberger, V. Mukhanov and T. Prokopec, Phys. Rev.
Lett. 69, 3606 (1992);

Phys. Rev. D48, 2443 (1993)

\item{48.}S. W. Hawking, Phys. Lett. B115, 295 (1982);

A. Linde, Phys. Lett. B116, 335 (1982);

A. A. Starobinski, Phys. Lett. B117, 175 (1982);

A. Guth and S. Y. Pi, Phys. Rev. Lett. 49, 1110 (1982);

J. M. Bardeen, P. S. Steinhardt and M. S. Turner, Phys. Rev. D28,

679 (1983)

\item{49.}G. Veneziano, Europhys. Lett. 2, 133 (1986)

\item{50.}A. Guth, Phys. Rev. D23, 347 (1981)

\item{51.}M. Gasperini, M. Giovannini, K. A. Meissner and G. Veneziano,
in preparation

\item{52.}F. Lucchin and S. Matarrese, Ref. 43

\item{53.}Y. Hu, M. S. Turner and E. J. Weinberg, Inflation as the
unique solution to the horizon and flatness problem,
Fermilab-Pub-92/363-A, astro-ph/9302002

\item{54.}T. Damour and A. M. Polyakov, The string dilaton and a least
coupling principle, IHES/P/94/1 (January 1994), hep-th/9401069

\item{55.}T. Banks, D. V. Kaplan and A. Nelson, Cosmological implications
of dynamical supersymmetry breaking, UCSD/PTH 93-26, RU-37,
hep-ph/9308292

\item{56.}M. Gasperini, Gen. Rel. Grav. 16, 1031 (1984)

\item{57.}J. Ellis, D. V. Nanopoulos and M. Quiros, Ref. 7

\item{58.}B. De Carlos et al., Ref.7

\item{59.}R. Brustein and G. Veneziano, The graceful exit problem in
string cosmology, CERN-TH.7179/94 (February 1994)

\vfill\eject

\centerline{\bf Figure captions}
\vskip 1 cm
\noi
\item{\bf Fig.1} The
dashed area defines the allowed windows for the dilaton
mass $m$ (given in units of $m_0=10^{-4}~eV$)
and the final inflation scale $H_1$ which are compatible
 with a present large contribution of non-relativistic dilatons
to $\Omega$, under the assumption that
they are produced with a fast enough growing spectrum,
$\da \geq 1$. For lower spectral slopes the allowed window is shifted
towards lower values of mass, according to eq.(1.9). Masses higher than
$100 \,MeV$ are excluded by dilaton decay,
masses lower than $10^{-4}\,eV$ by
tests of the equivalence principle. Inflation scales higher than $M_p$ are
excluded in order to avoid over-critical density
in the primordial relativistic particle production.

\vskip 1 cm

\item{\bf Fig.2} Maximum allowed region (inside the full lines) relative to
the cosmological production of dilatons with growing spectrum,
illustrated for three different spectral slopes $\da =0$, $\da=0.5$ and
$\da=1$ (the last case applies to all $\da\geq 1$).
The dilaton
mass is given in units of $m_0= 10^{-4} eV$. The lines marked by
$a,b,c,d,e,f,g$ represent the most significant bounds quoted in the
text, and
correspond respectively to: a) $m=m_0$, lower bound on
$m$ from the equivalence principle; b) $H_1=M_p$, upper bound on $H_1$
from the closure density; c) $T_r=1\, MeV$, lower bound on the reheating
temperature for nucleosynthesis; d) $m=M_p$, upper bound on $m>H_1$
from the closure density; e) $m=(H_2M_p^4H_1^{\da-4})^{1/(\da+1)}$,
upper bound on $m$ from the
present matter-to-radiation energy
density ratio; f) $m=(10^{-10}M_p^{-2}H_1^{4-\da})^{1/(2-\da)}$,
upper limit on entropy
production in dilaton decay from primordial baryogenesis; g)
$m=(10^{-2}M_p^{-2}H_1^{4-\da})^{1/(2-\da)}$,
upper limit on entropy production from
nucleosynthesis.

\end